\documentclass[manuscript]{aastex}

\shorttitle{Spectral Mapping of the Intermediate Polar DQ Herculis}
\shortauthors{Saito et al.}

\begin{document}

\title{Spectral Mapping of the Intermediate Polar DQ Herculis}

\author{R.~K.~Saito}
\affil{Departamento de Astronom\'{\i}a y Astrof\'{\i}sica, Pontificia
  Universidad Cat\'{o}lica de Chile, Vicu\~na Mackenna 4860, Casilla 306,
  Santiago 22, Chile}
\email{rsaito@astro.puc.cl}

\author{R.~Baptista}
\affil{Departamento de F\'{\i}sica, Universidade Federal de Santa
      Catarina, Trindade, 88040-900, Florian\'{o}polis, SC, Brazil}

\author{K.~Horne}
\affil{School of Physics and Astronomy, University of St. Andrews,
      KY16 9SS, Scotland, UK}

\and

\author{P.~Martell}
\affil{University of Wisconsin Center -- Marinette, 750 W. Bay Shore Street,
  Marinette, WI 54143, USA}

\begin{abstract}
We report an eclipse mapping study of the intermediate polar DQ~Her  based on
time-resolved optical spectroscopy ($\Delta\lambda \sim  3800-5000$\,\AA)
covering 4 eclipses. The spectra were sliced into 295 narrow passbands in the
continuum and in the lines, and the corresponding light curves were analysed
to solve for a set of monochromatic maps of the disk brightness distribution
and for the flux of an additional uneclipsed component in each band. 
Eclipse maps of the He\,II $\lambda 4686$ line indicate that an azimuthally-
and vertically-extended bright spot at disk rim is important source of
reprocessing of x-rays from the magnetic poles.   The disk spectrum is flat
with no Balmer or Helium lines in the inner regions, and shows double-peaked
emission lines in the intermediate and outer disk regions while the slope of
the continuum becomes progressively redder with increasing radius. The inferred
disk temperatures are in the range $T\simeq 13500 - 5000\,K$ and can be
reasonably well described by a steady-state disk with mass accretion rate of
$\dot{M}=(2.7\pm1.0)\times10^{-9}\,M_{\odot}\,yr^{-1}$.
A comparison of the radial intensity distribution for the Balmer lines reveals
a linear correlation between the slope of the distribution and the transition
energy. 
The spectrum of the uneclipsed light is dominated by Balmer and He\,I lines in
emission (probably from the extended nova shell) with narrow absorption cores
(likely from a collimated and optically thick wind from the accretion disk).
The observed narrow and redshifted Ca\,II $\lambda 3934$ absorption line in
the total light spectra plus the inverse P-Cygni profiles of the Balmer and
He\,II $\lambda 4686$ emission lines in spectra of the asymmetric component
indicate radial inflow of gas in the innermost disk regions and are best
explained in terms of magnetically-controlled accretion inside the white dwarf
magnetosphere. We infer projected radial inflow velocities of
$\sim200-500\,km\,s^{-1}$, significantly lower than both the rotational and the
free-fall velocities for the corresponding range of radii.  A combined net
emission He\,II plus H$\beta$ low-velocity eclipse map reveals a twisted
dipole emitting pattern near disk center. This is interpreted as being the
projection of accretion curtains onto the  orbital plane at two specific spin
phases, as a consequence of the selection in velocity provided by the spectral
eclipse mapping. 
\end{abstract}


\section{Introduction}

In many cataclysmic variable stars, the magnetic field of the white dwarf
(primary star) is sufficiently weak to be neglected. On the other hand, in
systems like the AM Her stars, the magnetic field dominates completely the
accretion flow \citep{2001cvs..book.....H}. Intermediate cases can lead to a
complex range of phenomena. This occurs in the intermediate polar stars,
where a magnetic field of moderate strength ($B \lesssim 10^{7}\,G$) allows the
combination of characteristics of non-magnetic systems (in the outer disk
regions) with characteristics of highly magnetic systems (in the regions close
to the white dwarf).

In intermediate polars the inner regions of the accretion disk are
disrupted by the magnetic field of the white dwarf and the infalling
plasma is forced to follow the field lines down to the star's surface
at one or both the magnetic poles, transforming kinetic energy into
radiation. Part of this energy can be reprocessed in other sites of the
binary, leading to pulsations at the white dwarf rotation period and/or
its beating with the orbital period \citep{1986ApSS.118..271W}.

DQ Her, or Nova Her 1934, is a deeply eclipsing cataclysmic variable (orbital
period $P=4.6\,h$) and the prototype of the subclass of intermediate polars,
showing a coherent $71\,s$ pulsation \citep{1956ApJ...123...68W}. Due to the
high inclination of the system \citep[$i=86.5\degr$;][]{1993ApJ...410..357H}
the white dwarf, accretion disk and bright spot are occulted by the secondary
star, allowing the emission from these different light sources to be separated
and spatially resolved by indirect imaging techniques such as eclipse-mapping
\citep{1985MNRAS.213..129H} and Doppler tomography
\citep{1988MNRAS.235..269M}.

Here we report the results of an eclipse-mapping study of the spectra and
structure of the accretion disk of DQ~Her based on time-resolved spectroscopy
of 4 eclipses collected on July 1987. The observations are described in
Section~\ref{obs}. Data analysis procedures are given in
Section~\ref{data_ana}. In Section~\ref{results} we investigate the disk
spatial structure in the emission lines and in the continuum and we present
spatially resolved spectra of the disk, gas stream and of the uneclipsed
light. The results are discussed in Section~\ref{discussion} and summarized in
Section~\ref{summ}. In the Appendix~\ref{appendixA} we discuss the ability to
reconstruct the full-width-half-maximum distribution of emission lines with
spectral mapping techniques.

\section{Observations}\label{obs}

\begin{figure*} 
\includegraphics[bb=0cm 0cm 19cm 28cm,angle=-90,scale=0.65]{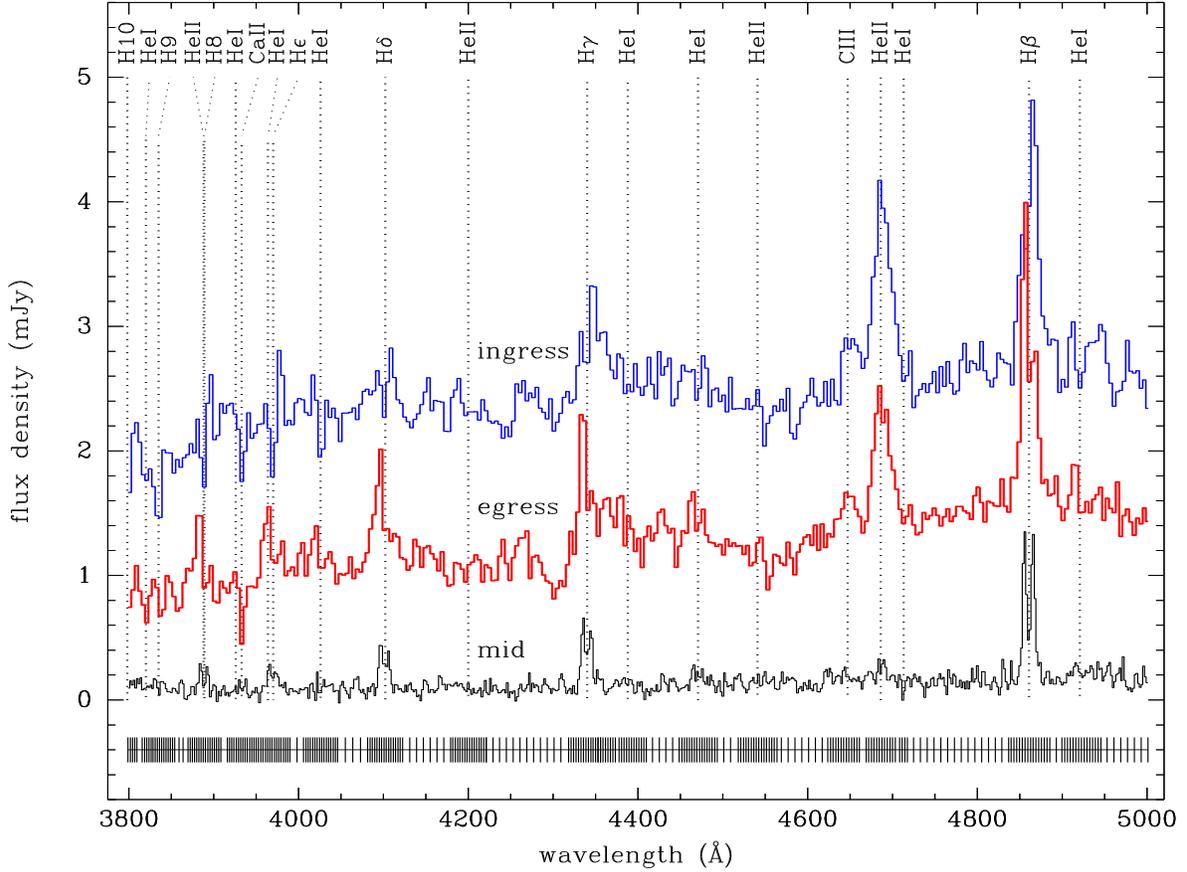}
 \caption{Average ingress (dark grey, phase range $-0.09$ to $-0.08$
cycle), egress (light grey, $+0.10$ to $+0.11$ cycle) and mid-eclipse
(black, $-0.02$ to $+0.02$ cycle) spectra of DQ Her, from the
combined eclipse cycles 62105, 62106, 62110 and 62111 (see text). Vertical
dotted lines indicate the emission/absorption lines. Horizontal ticks mark
the 295 passbands used to extract light curves. For a better visualization
the ingress and egress spectra are vertically displaced by $-1.4$ and
$1.7\,mJy$, respectively.}
\label{spec_med}
\end{figure*}

The data consist of 3272 spectra obtained with the Double Spectrograph at
the Hale 5\,m telescope, using the 2D-FRUTTI detector on the nights of
1987 July $3-4$. The runs cover the eclipse cycles 62105, 62106, 62110
and 62111, according to ephemeris of \cite{1981PASP...93..130A}. The
integration time for each spectrum was $10\,s$, and the usable spectral
coverage is $\Delta\lambda \sim 3800-5000$ \AA.  The reader is referred
to \cite{1995ApJ...448..380M} for a detailed description of the observations
and of the data reduction procedures.

The data were corrected to the rest frame of the white dwarf by removing its
orbital velocity $K_{1}=140~km\,s^{-1}$ \citep{1993ApJ...410..357H} from the
spectra, and then binned to a resolution of $130~km\,s^{-1}$ per bin
(2~\AA\,pixel\,$^{-1}$ at 4500 \AA). In the present work we analyse only the
data around eclipse, in the phase range $-0.1$ to $+0.1$, comprising a subset
of 1073 spectra from the original data set.

Fig.~\ref{spec_med} shows average spectra prior to (phase range $-0.09$
to $-0.08$ cycle), after ($+0.10$ to $+0.11$) and at mid-eclipse ($-0.02$ to
$+0.02$). The spectra show Balmer emission lines, a strong He\,II 4686
line, weak He\,I and He\,II lines and a few metal absorption lines on
top of a flat continuum. The emission lines show a clear double-peaked
profile at mid-eclipse. The Balmer lines are more prominent at mid-eclipse
than outside eclipse, indicating that their emission is less concentrated
towards disk center than that of the continuum.

\section{Data analysis}\label{data_ana}

\subsection{Light-curve construction}\label{lcurves}

The spectra were divided into 295 narrow passbands; 72 continuum passbands,
typically 8 \AA\ wide, and 223 velocity-resolved bands of $200~km\,s^{-1}$ for
the 22 lines marked in Fig.~\ref{spec_med}. In defining the velocity-resolved
line bins, we neglected the small systemic velocity of the object
\citep[$\gamma = 5 \pm 5~km\,s^{-1}$,][]{1993ApJ...410..357H}. For the
passbands including emission lines, the light curve comprises the total flux
at the corresponding bin, with no subtraction of a possible continuum
contribution.

Light curves were extracted for each band, combining the average flux
of the four data sets on the corresponding wavelength range in phase
bins of 0.0025 cycles. The error bars are taken as the standard
deviation with respect to the average flux at each phase bin. The
resulting light curves were phase-folded according to the ephemeris of
\cite{1981PASP...93..130A}.

\begin{equation}
\rm T_{mid}=HJD~2434954.943899(2) + 0.193620897(5)~E  
\hspace{2 pt}.
\vspace{12 pt}
\end{equation}

We remark that, because the 71\,$s$ oscillation was not removed from the
light curves, the resulting eclipse maps contain the combined light
distribution of the steady, unpulsed light and a time averaged distribution
of the pulsed light. Nevertheless, since the 71\,$s$ accounts for, at
most, $2 - 3$ \% of the total flux \citep{1995ApJ...448..380M}, the influence
of the pulsed light on the overall disk brightness distribution is small. 
Light curves and eclipse maps for the H$\beta$ and He\,II 4686 lines are
shown in Figs.~\ref{hbet_curves} and \ref{heii_curves}, respectively. 
Contour curves in these figures enclose the regions of the eclipse maps
at and above the 15$\,\sigma$ level of statistical significance.

\subsection{Eclipse-mapping} \label{mem}

We used maximum-entropy eclipse-mapping techniques \citep{1985MNRAS.213..129H,
1993AA...277..331B} to solve for a map of the disk brightness distribution and
for the flux of an additional uneclipsed component in each spectral band. The
reader is referred to \cite{2001LNP...573..307B} for a review on the
eclipse-mapping method.

As our eclipse map we adopted a flat grid of $65\times65$ pixels centered on
the primary star with side $2\,R_{L1}$, where $R_{L1}$ is the distance from
the disk center to the inner Lagrangian point ($L_{1}$). In this work we
adopted a Roche lobe size of $R_{L1}=0.766\,R_{\odot}$
\citep{1980ApJ...241..247P}. The eclipse geometry is defined by the mass ratio
$q$, which sets the shape and the relative size of the Roche lobes, and the
inclination $i$, which determines the shape and extension of the shadow of the
secondary star projected onto the orbital plane. We adopted the values by
\cite{1993ApJ...410..357H}, $q=0.66 \pm 0.04$ and
$i=86.5\degr \pm 1.6\degr$. This combination of parameters ensures that the
white dwarf is at the center of the map.

Before applying eclipse mapping techniques to a high inclination system such
as DQ~Her, it is important to address the question whether the disk rim can
hide parts of the accretion disk from view (thereby leading to  unreliable
results). In particular, a vertically-extended, thick disk rim could lead to
an important self-obscuration effect on the disk side closest to the observer
if the disk half-opening angle is $\beta \gtrsim (90\degr-i)$.  An estimate of
the disk half-opening angle in DQ~Her comes from the recent work of
\cite{2009ApJ...693L..16S}.  They applied eclipse mapping techniques to trace
the changes in surface brightness distribution along the 71\,$s$ pulsation
cycle and found that a vertically-extended bright spot at disk rim is the main
site of emission of the $71\,s$ optical oscillations. The upper side of the
bright spot is illuminated by one of the magnetic poles once per cycle, while
the lower side of the bright spot is illuminated by the other pole roughly
half a cycle later. From the difference in the projected position of the upper
and lower bright spot sides in their eclipse maps, they inferred a vertical
thickness of $H= (7.4\pm 0.4)\times 10^{-3}\, R_\odot\, [\tan(i)/16.35]^{-1}$
\footnote{In \cite{2009ApJ...693L..16S} this value is mistakenly quoted as the
bright spot vertical half-thickness.}, where the term within brackets is unity
for $i=86.5\degr$. Combined with a bright spot radial distance of $R_{BS}=
(0.44\pm 0.02)\,R_\odot$ (see Section~\ref{structures}), this leads to $\beta=
\arctan(H/2R_{BS})= 0.5\degr - 0.002\degr$ for $i=86.5\degr-89\degr$. The
lowest range of values for $\beta$ is hard to be interpreted by the
steady-state disks model, where flared disks are expected, with the vertical
tickness $H$ proportional to the mass accretion rate $\dot{M}$ and increasing
with radius with by factor $H\propto r^{9/8}$. The inferred value of $\beta$
decreases with increasing inclination, and the inequality $\beta <
(90\degr-i)$ holds for any reasonable inclination ($i>70\degr$). Hence,
self-obscuration by the disk rim is not an issue in DQ~Her.  The fact that the
bright illumination pattern that rotates with the $71\,s$ pulsation period in
the inner disk regions remains visible when it is passing in the disk side
near the observer \citep{2009ApJ...693L..16S} yields additional support for
the above statement. We are therefore confident that no distortion effects
caused by self-obscuration by the disk rim arise when applying eclipse mapping
techniques to the DQ~Her light curves.

For the reconstructions we used the default of limited azimuthal smearing of
\cite{1992AA...265..159R}, which is better suited for recovering asymmetric
structures than the original default of full azimuthal smearing
\citep[c.f.,][]{1996MNRAS.282...99B}. Simulations showing the ability of the
eclipse mapping method to reconstruct asymmetric brightness distributions
(even under unfavorable conditions of low S/N and incomplete phase coverage)
are presented and discussed by \cite{2001LNP...573..307B} and
\cite{2001MNRAS.324..599B}. We used a radial blur width $\Delta r = 0.0185
R_{L1}$ and an azimuthal blur width $\Delta \theta = 20 \degr$.

\begin{figure}
\includegraphics[bb=-5cm 1cm 10cm 24.2cm,scale=.60]{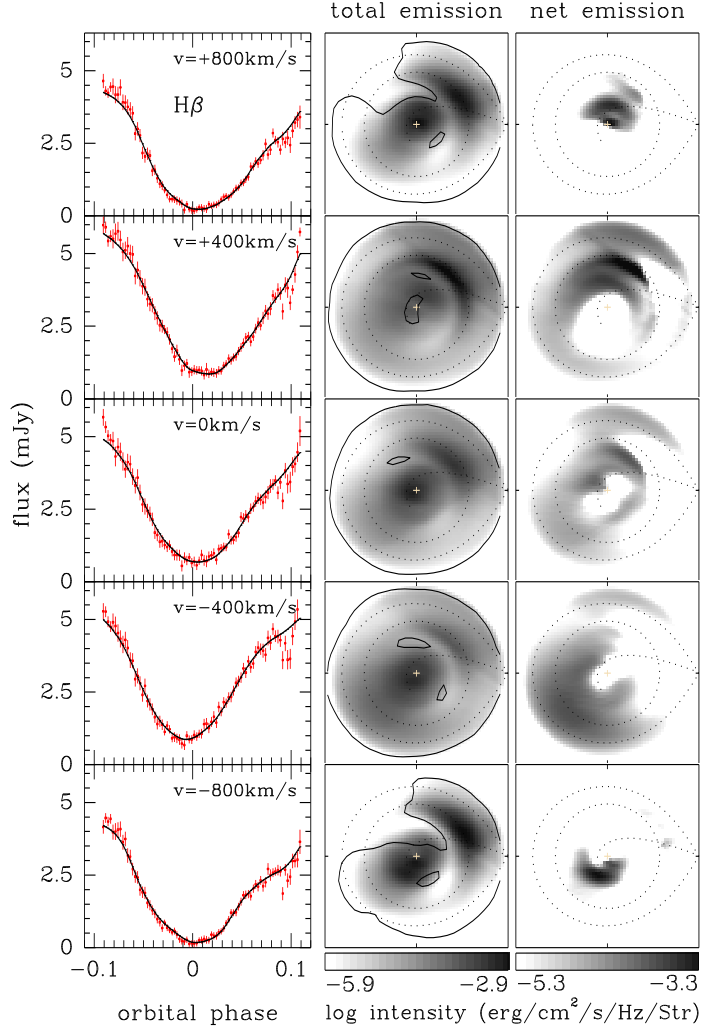}
 \caption{Data and model light curves and eclipse maps for H$\beta$
($\lambda 4861$) at five velocity bins ($v = \pm 800$, $\pm 400$, $0~km\,
s^{-1}$, $\Delta v=200~km\,s^{-1}$) . The left panels show the data (dots
with error bars) and model (solid lines) light curves. The middle panels
show the corresponding eclipse maps for the total emission (line +
continuum) in a logarithmic greyscale.  Brighter regions are indicated
in black; fainter regions in white.  A cross marks the center of the disk;
dotted lines show the Roche lobe, the gas stream trajectory and a circle
of radius R$_{BS}= 0.57\,R_{L1}$ corresponding to the bright spot position;
the secondary is to the right of each map and the stars rotate
counter-clockwise. Contour curves enclose the regions at and above the
15$\,\sigma$ level of statistical significance.  The right panels show
eclipse maps of the net line emission in each case; the notation is
similar to that of the middle panels. The horizontal bars indicate the
logarithmic intensity level of the greyscale in each column.}
\label{hbet_curves}
\end{figure}

\begin{figure}
\includegraphics[bb=-5cm 1cm 10cm 24.2cm,scale=.60]{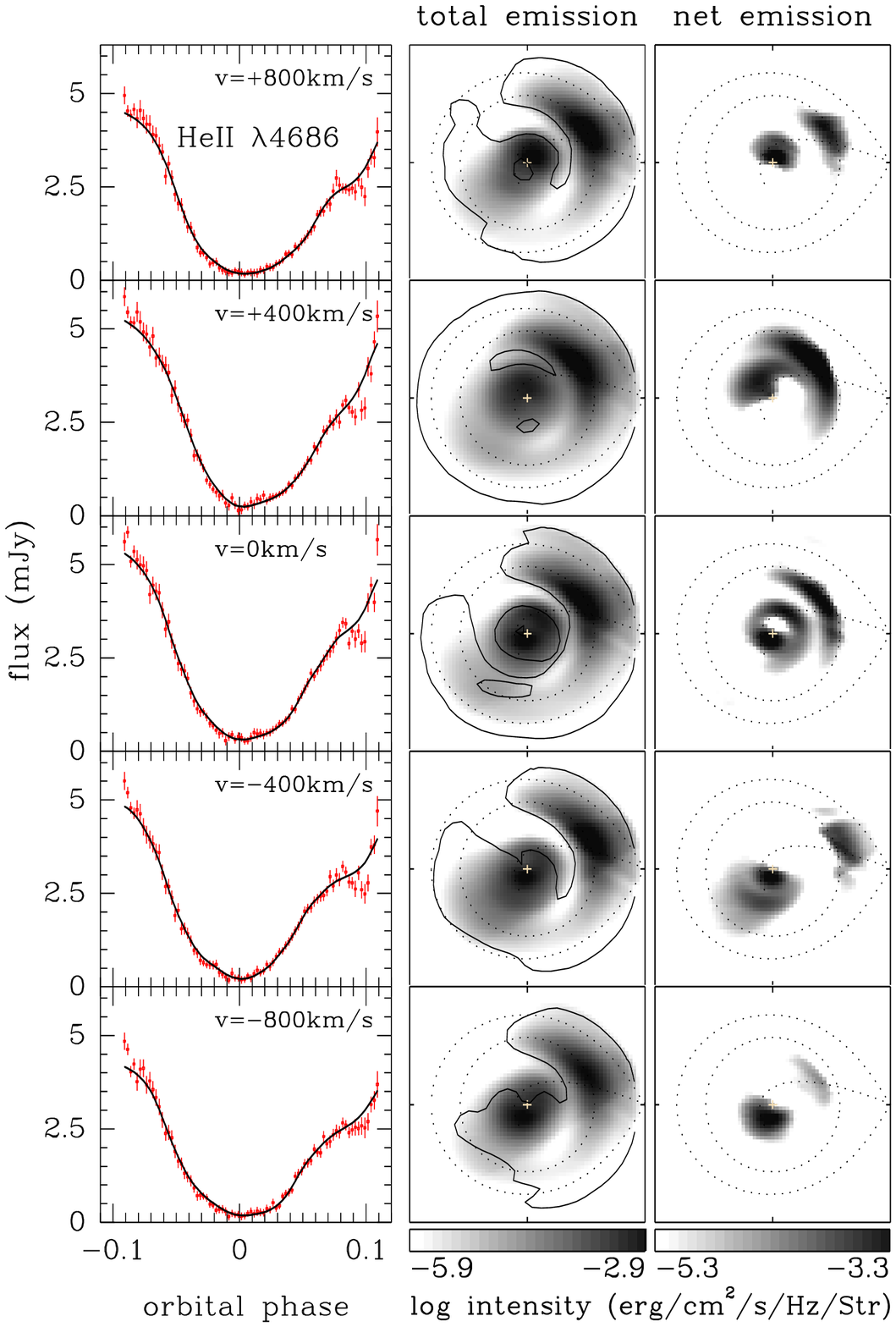}
 \caption{Velocity-resolved light curves (left) and eclipse maps
(middle and right) for He\,II 4686. The notation is similar to that of
Fig.~\ref{hbet_curves}.}  \label{heii_curves}
\end{figure}

The statistical uncertainties in the eclipse maps were estimated with a
Monte Carlo procedure \citep[e.g.,][]{1992AA...265..159R}. For each narrow band
light curve, a set of 20 artificial light curves was generated in which
the data points were independently and randomly varied according to a
Gaussian distribution with standard deviation equal to the uncertainty at
that point. The light curves were fitted with the eclipse-mapping code to
produce a set of randomized eclipse maps. These were combined to produce
an average map and a map of the residuals with respect to the average,
which yields the statistical uncertainty at each pixel. The uncertainties
obtained with this procedure were used to estimate the errors in the
derived radial intensity and temperature distributions as well as in
the spatially resolved spectra. We also divided each eclipse map by the
map of the residuals with respect to the average to produce maps of the
inverse of the relative errors, or signal-to-noise ratio maps
\citep{2004AJ....128..411B}. The S/N maps are overplotted on the corresponding
eclipse maps in Figs.~\ref{hbet_curves} and \ref{heii_curves} as
a contour line for $S/N=15$. All structures in the eclipse maps are
significant at a confidence level equal to or above $15\sigma$.

\section{Results}\label{results}

\subsection{Disk structures} \label{structures}

Fig.~\ref{hbet_curves} shows eclipse maps of H{$\beta$} for a set of
5 velocity bins. Similarly, Fig.~\ref{heii_curves} shows
velocity-resolved eclipse maps for the He\,II 4686 line. The
velocity-resolved H{$\beta$} (and, to a less extent, the He\,II 4686
line) light curves show the expected behavior for the eclipse of gas
rotating in the prograde sense (rotational disturbance), with the blue
side being eclipsed earlier than the red side. The eclipse maps in
symmetric velocity bins do not show the reflection symmetry around the
line joining both stars expected for the emission from a symmetric
disk around the white dwarf.

\begin{figure}
\includegraphics[bb=1cm -15cm 20cm 10cm,angle=-90,scale=.3]{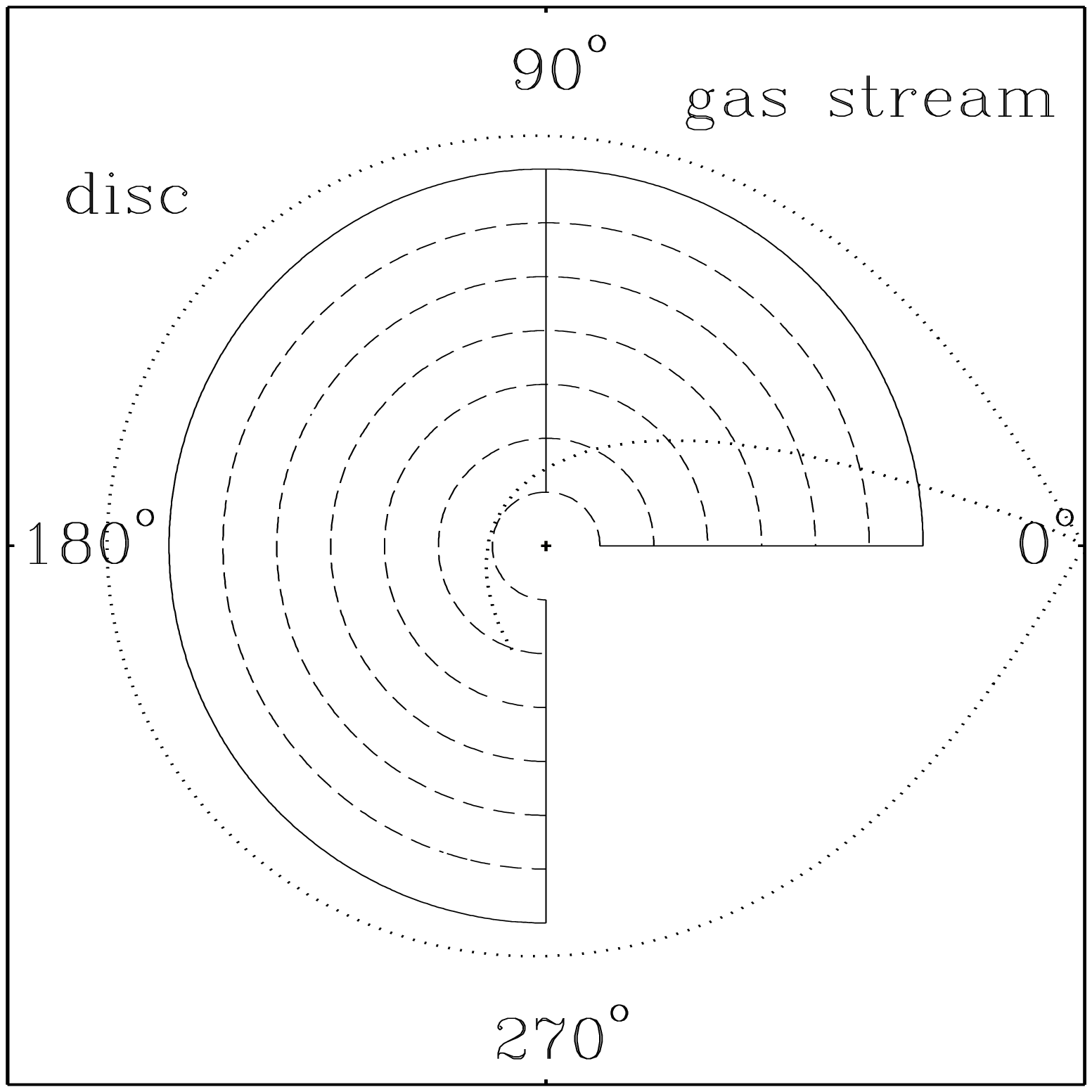}
 \caption{Schematic diagram showing the regions defined as ``disk''
and ``gas stream''. Dashed lines mark the annular regions of width 0.1
R$_{L1}$ used to extract spatially resolved spectra. Dotted lines show
the projection of the primary Roche lobe onto the orbital plane and
the gas stream trajectory. Azimuths are measured with respect to the
line joining both stars and increase counter-clockwise. Four reference
azimuths are labeled in the figure.}\label{disk_esq}
\end{figure}

The clear egress shoulder in the light curves reveals an
azimuthally-extended asymmetry in the outer parts of the disk extending
$\simeq 90 \degr$ ahead of the intersection of the stream trajectory with
the disk rim. This azimuthally-extended bright spot indicates that the
shock of the inflowing gas with the outer disk is radiatively inefficient,
with the emission pattern spreading along the direction of gas rotation.
Moreover, if the bright spot cannot cool efficiently it should also
expand in the vertical direction, leading to a vertically-extended bulge
in the outer disk that becomes a natural site for reprocessing of x-ray
radiation from the rotating magnetic poles -- in line with the findings
of \cite{2009ApJ...693L..16S}.  From the radial position of the maximum
intensity of the bright spot, we estimate an outer disk radius of
R$_{BS} = 0.57 \pm 0.03\, R_{L1}$. 

In order to emphasize the structures observed in the line maps, we calculated
net line emission maps. These were obtained by combining continuum eclipse
maps on the short and long wavelength sides of the target emission line and by
subtracting the derived average continuum map from each of the velocity
resolved line maps.  Because of our choice to display the maps in a
logarithmic grayscale (that only handles positive intensities), regions where
the resulting intensities are negative (indicating line absorption) were
suppressed from the net line emission maps shown in Figs.~\ref{hbet_curves}
and \ref{heii_curves}.  The Balmer lines net emission maps show a dearth of
emission at the bright spot position, indicating that these lines are in
absorption at the infalling stream impact site (see Section~\ref{spectra}). In
contrast, the He\,II 4686 line appears clearly in emission at the bright spot
position (Fig.~\ref{heii_curves}, right-hand panels). The emission is stronger
in the positive velocity maps but weak blueshifted emission is also seen in
the $-400$ and $-800~km\,s^{-1}$ velocity maps. The latter is likely the
result of gas splashing away from the accretion disk (with a velocity
component towards the observer) after the collision of the infalling gas
stream with the disk rim. The redshifted emission reflects the fact that disk
gas at the bright spot position is mostly moving towards the far side of the
disk and, therefore, away from the observer.
 
Because the temperatures in the outer disk regions ($T \simeq 6000-7000 \,K$,
see Section~\ref{temperatures}) are not enough to power the high-excitation
He\,II 4686 line, the strong He\,II emission from this region is an indication
that the bright spot must be an important source of reprocessing of x-ray
radiation from the magnetic poles/curtains -- again in line with the findings
of \cite{2009ApJ...693L..16S}.  This is consistent with and leads to a natural
explanation for the results of \cite{1978ApJ...224..570P} --- who
found that the amplitude of the $71\,s$ pulsations is variable along the orbit
and reaches maximum around phase $\phi \sim 0.25$ --- and of
\cite{1985...conf...98} --- who correctly advanced the conclusion that the
orbital behavior of the $71\,s$ pulsations amplitude requires that the x-ray
reprocessing site be a disk rim with variable thickness and maximum visibility
at phases $0.2-0.3$. This coincides with the phase range where the upper side of
the azimuthally- and vertically-extended bright spot is best seen.

The zero-velocity H{$\beta$} net line emission map shows a peculiar
``heart''-shaped distribution close to disk center ($R< 0.3\,R_{L1}$)
with major axis aligned at an angle of $\simeq 150\degr$ with respect
to the line joining both stars (Fig.~\ref{hbet_curves}).
Morphologically similar structures are also seen in the $\pm 200$ and
$\pm 400\,~km\,s^{-1}$ velocity maps, with the positive velocity maps
showing the receding part of the structure (i.e., the upper
hemisphere of the eclipse maps in Figs.~\ref{hbet_curves} and 
\ref{heii_curves}) and the negative velocity maps showing the
leading part of the structure (i.e., the lower hemisphere of the
eclipse maps in Figs.~\ref{hbet_curves}~and~\ref{heii_curves}).
An asymmetric structure resembling a dipole pattern and aligned at a
similar orientation can be seen in the zero-velocity He\,II 4686 net
emission map (Fig.~\ref{heii_curves}).
We will return to this point in Section~\ref{curtains}.

\subsection{Spatially resolved spectra} \label{spectra}

Motivated by the distinct emission observed at the bright spot and gas
stream regions and in order to avoid contamination by bright spot 
emission that could affect the spectrum of the near side of the disk
(the hemisphere closest to the secondary star in the eclipse maps of
Figs.~\ref{hbet_curves}~and~\ref{heii_curves}), we divided the eclipse
maps in two distinct regions: ``disk'' and ``gas stream''.
We define the ``disk'' as the disk section between azimuths $90\degr$
and $270\degr$, and ``gas stream'' as the region between azimuths
$0\degr$ and $90\degr$ (see Fig.~\ref{disk_esq}).
We further defined annular sections of width $0.1~R_{L1}$ to extract
spatially-resolved spectra as a function of radius. In order to isolate
the contribution of possible asymmetric brightness sources in the eclipse
maps (e.g., the bright spot, gas stream or accretion curtains) we separated
the total emission into its symmetric and asymmetric disk emission
components. The symmetric component is obtained by slicing the disk into
a set of radial bins, computing the average of the lower half of the
intensities in each bin, and fitting a smooth spline function to the
resulting set. The spline fitted intensity in each annular section is
taken as the symmetric component. The asymmetric component is obtained by
subtracting the symmetric component from the original eclipse map. This
procedure essentially preserves the baseline of the radial profile,
removing all azimuthal structure from the symmetric component. The
statistical uncertainties affecting the fitted intensities are estimated
with a Monte Carlo procedure (Section~\ref{mem}).

\begin{figure}
\includegraphics[bb=-4cm 1cm 18cm 23cm,scale=.60]{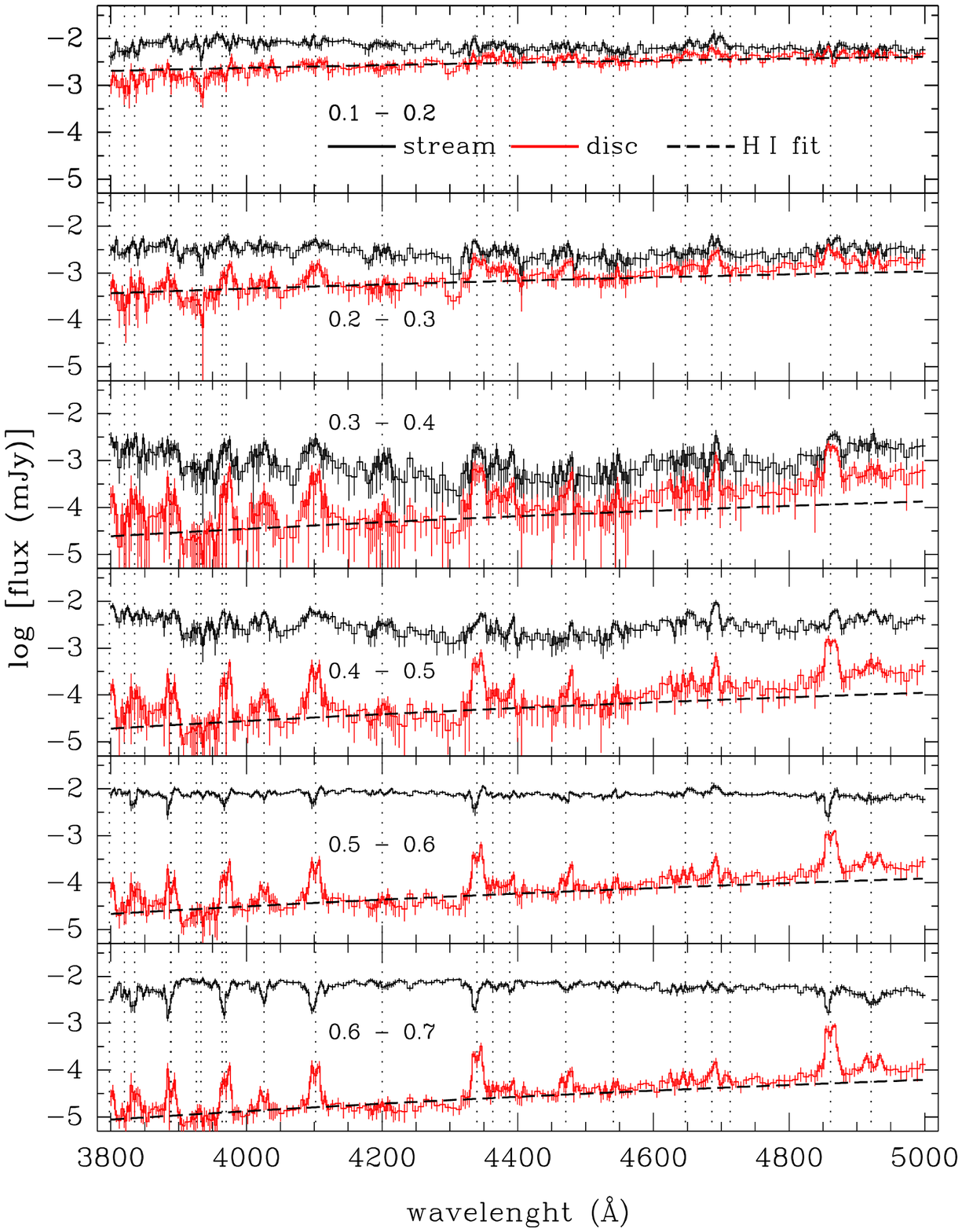}
 \caption{Spatially resolved spectra for the sections defined as
``disk'' (90$\degr$ -- 270$\degr$, light grey) and ``gas stream''
(0$\degr$ -- 90$\degr$, dark grey) for a set of six annular
regions (labels in units of $R_{L1}$). Dashed lines show the best-fit
H\,I model for the disk spectra. Vertical dotted lines mark the
major transition lines of the spectra.}\label{spec_disk}
\end{figure}

Fig.~\ref{spec_disk} shows spatially resolved spectra for the symmetric
disk component and for the ''gas stream'' region (total emission). 
In the inner regions ($R\lesssim 0.2\,R_{L1}$) the disk spectrum shows a
flat and featureless continuum with no Balmer or He\,II lines. The disk
spectrum becomes progressively redder and fainter with increasing radius,
indicating the existence of a radial temperature gradient. The Balmer
and He\,II lines start to appear in emission at intermediate disk regions
($R\simeq 0.3\,R_{L1}$) and become stronger with increasing radii.
The spectrum of the ''gas stream'' is systematically bluer and brighter
than the disk spectrum at the same radius and the difference increases
with radius.  In contrast to the observer in the disk spectra, in the
outer regions ($R>0.5\,R_{L1}$) the lines appear in absorption, indicating
that the emitting gas is optically thick. The P~Cygni profile of the 
Balmer lines in the ``gas stream'' spectrum framing the bright spot, at
$R= (0.5-0.6)\,R_{L1}$, is an indication that part of the colliding gas
stream deflects back towards the observer, in agreement with the inference
previously drawn from the He\,II $\lambda 4686$ net emission line maps.

The lack of Balmer lines in the innermost disk regions may be interpreted in a
scenario of magnetically-controlled accretion for radii smaller than the
Alfv\'en radius. Assuming for DQ Her a white dwarf mass of
$M_{1}=0.6\,M_{\odot}$ \citep{1995ApJ...454..447Z}, an accretion rate of
$\dot{M}=2.7\times10^{-9}\,M_{\odot}\,yr^{-1}$ (Section~\ref{temperatures})
and a white dwarf magnetic moment in the range $\mu = 10^{32} -
10^{33}\,G\,cm^{3}$ \citep[typical for intermediate
polars,][]{1994PASP..106..209P}, we estimate an Alfv\'en radius for DQ Her in
the range $R_{A} \simeq (0.05 - 0.25)\,R_{L1}$. It is expected that inside
this radius the disk will be disrupted, with the gas being channelled by the
magnetic field lines onto accretion curtains down to the magnetic poles at the
white dwarf surface. This produces magneto-{\it bremsstrahlung} (free-free)
radiation by the optically thin gas, which leads to a spectrum basically flat
for the spectral range between the self-absorption limit and a
thermal-cutoff. Hence, the spectrum observed in the innermost disk regions
(i.e., flat and featureless) could plausibly be interpreted in this context
\citep{1959ApJ...130..110K, 1988prco.book..199W}.

A relatively narrow (FWHM $\lesssim 400\,~km\,s^{-1}$) and redshifted  (by
$\simeq 200\,~km\,s^{-1}$) Ca\,II $\lambda 3934$ absorption line can be seen
both in the disk and in the gas stream spectra up to $R\sim
0.3\,R_{L1}$. (There are hints that a similarly narrow and redshifted Ca\,II
$\lambda 3968$ absorption line is also present in the disk and gas stream
spectra at the same radial range, although we assign a low weight to this
statement as this line is heavily blended with the broad $H\epsilon$ emission
line.)  This corresponds to the Ca\,II s-wave seen in absorption in the
trailed spectrogram of the same data set and which \cite{1995ApJ...448..380M}
associated to the accretion disk. 

Could the Ca\,II line be the result of absorption of disk radiation
by a cool and vertically-extended disk rim along the line of sight?
The evidences say no. If this was the case, the line strength would
increase with radius for spectra of the near side of the disk since the
obscuration effect increases as one moves towards the disk rim in the
near side. On the contrary, the observed Ca\,II line strength decreases
with increasing radius in the spectra of the ``gas stream'' 
(Fig.~\ref{spec_disk}) as well as in the symmetric component of spectra
extracted for the near side of the disk (not shown here).

The temperatures inferred for the inner disk regions ($T > 10^4\,K$,
see Section~\ref{temperatures}) are too high to allow for significant
Ca\,II emission. Thus, the Ca\,II absorption should be produced in
intervening cooler gas along the line of sight. Interesting inferences
can be drawn from the dynamics of this intervening gas. Gas rotating in 
Keplerian orbits at this distance from the white dwarf should lead to
broad (FWHM $\sim 1000-3000\,~km\,s^{-1}$) and double-peaked absorption
lines with no blueshift/redshift. 
The observed narrow, single-peaked and redshifted Ca\,II line indicates
that the gas dynamics in the absorbing region is dominated by radial
motion (in opposition to azimuthal, Keplerian motion). Depending on where
the absorption occurs along the line of sight, we may be witnessing gas
outflow (if absorption occurs closer to the orbital plane, at the far
side of the disk) or gas inflow (if absorption occurs closer to the
observer along the line of sight, at the near side of the disk).

In the radial outflow scenario DQ~Her would be acting as a propeller
\citep[e.g.,][]{1997MNRAS.286..436W}, expelling gas from inside the
magnetosphere towards larger radii. The corotation radius for DQ~Her
($R_{co}= 0.031\,R_\odot = 0.041\,R_{L1}$) is comparable to (and even smaller
than) our lower estimate of the Alfv\'en radius, suggesting that  plasma blobs
that attach to field lines at $R_{co}<R<R_A$ should be centrifugally
accelerated \citep[][Chapter~7]{1995cvs..book.....W}. Accordingly, one would
expect to see blueshifted Ca\,II absorption in spectra of the near side of the
disk in account of gas expelled towards the observer.  However, contrary to
this expectation, the Ca\,II absorption line is also redshifted both in the
``gas stream'' spectra and in the symmetric component of spectra extracted for
the near side of the disk. Additional problems with this scenario include the
difficulties to explain why the outflow velocity would be so low
($v_{out}\simeq 200\,km\,s^{-1}$), why it does not change with radius, and why
no Ca\,II absorption is seen in the outer disk regions.

The radial inflow alternative leads to a magnetically-controlled accretion
scenario, in which the Ca\,II absorption line is produced when relatively
cool regions of the accretion curtain are seen projected against hot gas
in the background.  Inside the magnetosphere ($R < R_A$) two effects
contribute to broaden the emission/absorption lines: (i) the gas corotates
with the white dwarf magnetic field, the spin velocity of which increases
with radius, $v_\phi= R\,\omega_{\rm spin}$\,; (ii) at the same time,
gas moves inwards along field lines with a radial velocity $v_r\leq v_{ff}
\propto R^{-1/2}$. Because DQ~Her is a fast rotator, large azimuthal
velocities ($v_\phi \simeq 2300 - 12000\,km\,s^{-1}$) are expected for the
range of radii of interest, $R= 0.05 - 0.25 \,R_{L1}$, and lines produced
in this region should be quite broad. Then, why is the Ca\,II line so
narrow?

At this point it is worth noting that because an eclipse map yields a snapshot
of the eclipsed brightness distribution over a time scale much longer than the
$71\,s$ pulsation period, the contribution of the rapidly rotating
magnetosphere to our spatially-resolved spectra is a time averaged spectrum
over all orientations (or, pulse phases).  Adopting the oblique dipole
accretor model of \cite[][see their Fig.~3]{2009ApJ...693L..16S}, significant
absorption should occur around spin phase $\phi_{\rm spin}=0.5$ --- when the
upper accretion curtain covers the largest portion of the hot inner
magnetosphere and/or the (also hot) innermost regions of the far side of the
accretion disk. However, at  this orientation the spin component of the
accretion curtain velocity field is mostly in the plane of the sky and the
projected velocity along the line of sight is essencially $v_r$. Note that, at
quadrature ($\phi_{\rm spin}= 0.25$ and 0.75), the absorption of radiation by
the upper accretion curtain is not only significantly reduced but is also
smeared in wavelength because of the contribution of a large spin velocity
component to the line profile.  Thus, the observed absorption line profile is
dominated by the spectral contribution of the accretion curtains at $\phi_{\rm
spin}=0.5$.  The magnetically-controled accretion scenario implies radial
inflow velocities of $v_r \sim 200\,km\,s^{-1}$, significantly lower than both
the rotational and the free-fall velocities for that range of radii.  This is
line with inferences drawn by \cite{1991MNRAS.252..386W}, who pointed out
that, for rapidly rotating intermediate polars, the accretion shock may occur
almost at $R_A$ and the infall from $R_A$ to the white dwarf surface is
unlikely to have a significant radial component.

\subsection{The emission lines} \label{emission}

\begin{figure*}
\includegraphics[bb=1cm -1cm 19cm 28cm,angle=-90,scale=0.58]{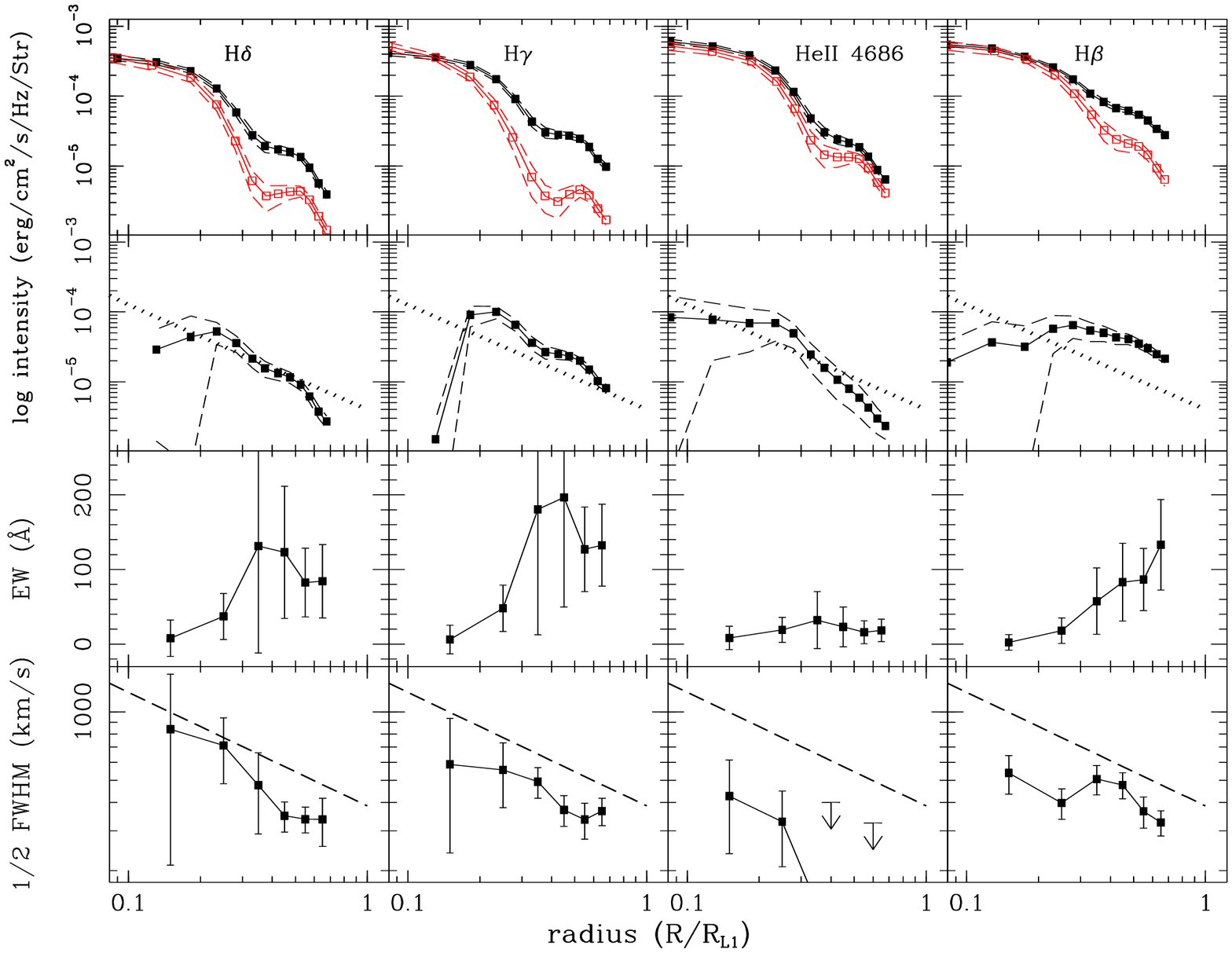}
 \caption{Top: Radial intensity distributions of selected optical
lines (filled symbols) and of the corresponding adjacent continuum
(open symbols). Dashed lines indicate the 1$\,\sigma$ limit in each
case. Second row from the top: Net line emission radial
distributions. Dotted lines show the $I \propto r^{-1.5}$ radial
dependence, derived by \cite{1990ApJ...364..637M}. Third from top: EW as a
function of radius. Bottom: FWHM of the lines as a function of radius.
Dashed lines show the law $v \propto r^{-1/2}$ expected for gas rotating
in Keplerian orbits around a white dwarf of mass $M_{1}=0.6\,M_{\odot}$.
The predicted spin velocity of material channelled by the white dwarf
magnetic field is too high to appear in the velocity range plotted in
the panels (see text).}
\label{ew}
\end{figure*}

In this section we analyse the radial behavior of Balmer (H$\beta$,
H$\gamma$ and H$\delta$) as well as the He\,II 4686 lines. Fig.~\ref{ew}
shows the radial intensity distribution for the lines and for the adjacent
continuum (top), the radial intensity distribution for the net line
emission (second row from top), the radial run of the equivalent width
(EW, third row from top) and the radial run of the full width at
half-maximum (FWHM, bottom). The diagrams were derived from the symmetric
disk component (Section~\ref{spectra}). The performance of the eclipse mapping
technique to reconstruct FWHM distributions is discussed in the
Appendix~\ref{appendixA}.

The intensities in the line maps are higher than in the adjacent continuum
at intermediate and outer disk regions, indicating that the lines appear 
in emission there. The Balmer lines disappear in the continuum in the
innermost disk regions ($R< 0.2\,R_{L1}$), and the corresponding net line
emission becomes unreliable or meaningless. For H$\gamma$ and H$\delta$
the line intensity is lower than the continuum level at disk center
indicating that these lines are in absorption there. These regions are
not plotted in the corresponding net emission panels.

The net line emission decreases in strength with increasing radius in
the intermediate and outer disk regions. The radial dependence has a
particular slope for each line, indicating that the empirical law
$I \propto r^{-1.5}$ \citep{1990ApJ...364..637M} is not adequate to describe the
behavior of the emission lines in DQ~Her. Moreover, we found that, for
the Balmer lines, the radial dependence is linearly correlated with the
energy required to excite the line. This result will be discussed in
Section~\ref{energy-slope}.

The EW of the Balmer lines increase with radius with maxima of $EW \simeq 100
- 400$\,\AA\ at $R= 0.5\,R_{L1}$. In contrast, the He\,II 4686 line shows an
$EW \simeq 20$\,\AA, roughly independent of radius. In all cases, the EW
becomes negligible (or negative) at the innermost disk regions because the
continuum intensities reach (or exceed) the line intensities.

In the intermediate and outer disk regions ($R \geq 0.3\,R_{L1}$) the
slope of the observed FWHM distribution of the Balmer lines is fairly
consistent with the expected $v_{K} \propto R^{-1/2}$ law for gas
rotating in Keplerian orbits. However, the observed FWHM values are
systematically lower than the predicted for a disk around a white dwarf
with $M_{1}=0.6\,M_{\odot}$  (the dashed line in the lowermost panels of
Fig.~\ref{ew}) by $\sim$ 30-40\%, suggesting that either the adopted
white dwarf mass is overestimated or that the emitting gas has
sub-Keplerian velocities. No reasonable error in the estimated inclination
could account for this discrepancy.  Moreover, in order to reconcile the
Keplerian velocity distribution with the observed FWHM values, an
unrealistically low-mass white dwarf ($\leq 0.35\,M_\odot$) would be
needed. It is therefore likely that the disk gas has sub-Keplerian
velocities. The He\,II 4686 FWHM is hard to measure and becomes unreliable
for $R \gtrsim 0.3\,R_{L1}$, because the net line intensity is decreasing
sharply with radius and the EW is relatively small.

\begin{figure*}
\includegraphics[bb=2cm 0cm 19cm 28cm,angle=-90,scale=0.60]{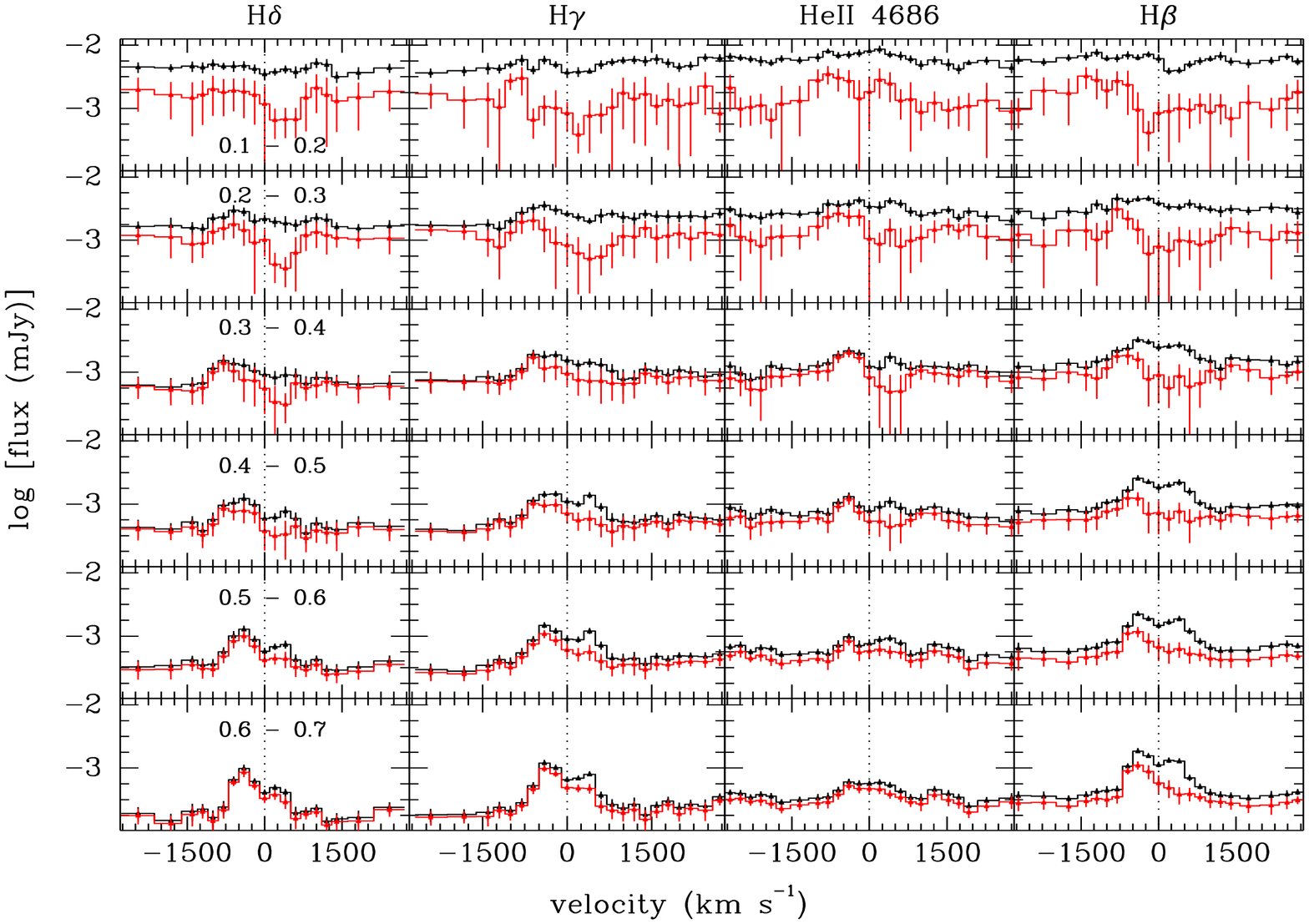}
 \caption{The velocity resolved spectra for the four most prominent
emission lines in the DQ Her spectra comparing the total emission (black)
and the asymmetric component emission (grey). Vertical dotted lines mark
the line center in each case}\label{spec_veloc}
\end{figure*}

Fig.~\ref{spec_veloc} shows the velocity profile for the total and the
asymmetric component of the emission lines. In the total emission spectra
the lines appear in emission in the intermediate and outer regions
($R\gtrsim0.2\,R_{L1}$) becoming progressively sharper (stronger and
narrower) with increasing radius. The Balmer and the He\,II 4686 lines
show a clear double-peaked profile with the blue peak generally stronger
than the red peak in the Balmer lines. The He\,II line changes from a
double-peaked profile at small radii to a single-peaked profile in the
outer disk regions.

The differences in behavior between the Balmer lines and the He\,II 4686
line has already been observed in previous analyses of DQ Her. The $71\,s$
pulsation, evident in He\,II, is absent \citep{1978ApJ...226..963C} or
negligible \citep{1995ApJ...448..380M} in the Balmer lines. This result
reflects the different ionization/excitation conditions between hydrogen
and heavier ions.

The spectra of the asymmetric component evince inverse P-Cygni profiles
for all lines, with a 'red' absorption component on top of a 'blue' 
emission peak. The strength of the absorption component decreases
with increasing radius and disappears for $R > 0.4\,R_{L1}$. 
The $H\beta$ absorption component is narrow and slightly blueshifted
(by $\simeq -200\,km\,s^{-1}$) in the innermost regions, and 
progressively broadens and moves towards positive velocities with
increasing radius, reaching trough velocities of $\simeq +500\,km\,s^{-1}$
at $R=(0.3-0.4)\,R_{L1}$. The absorption component is always redshifted
for the other lines, also with velocities of $\simeq +500\,km\,s^{-1}$.
The emission components are centered at $\simeq -500\,km\,s^{-1}$.
The blueshifted emission seen in spectra of the asymmetric component
of the far side of the disk indicates that emitting gas of a non
axi-symmetric brightness source is falling radially towards the white
dwarf at disk center. Consistently, absorbing gas along the line of
sight (possibly at the near side of the disk) is also falling radially
towards the white dwarf at disk center with similar radial inflow
velocities ($v_r\sim 500\,km\,s^{-1}$). These evidences are in agreement
with those inferred from the behavior of the Ca\,II $\lambda 3934$
absorption line; they give additional support to the
magnetically-controlled accretion scenario put forward in the
preceeding section.

\subsection{The uneclipsed component} \label{uneclipsed}

At the high inclination of DQ~Her, there is no region in the far side of the
disk which is never eclipsed and, therefore, it is not possible to dump extra
flux at mid-eclipse without affecting the shape of the model eclipse light
curve. Any extra flux at mid-eclipse is unambiguously attributed to
'uneclipsed light', i.e., the fraction of the total light which cannot be
included in the eclipse map. Thus, the uneclipsed component in DQ~Her arises
from light outside of its orbital plane (or from the outward-facing hemisphere
of the eclipsing secondary star).

\begin{figure}
\includegraphics[bb=3.5cm -6cm 19cm 18cm,angle=-90,scale=0.4]{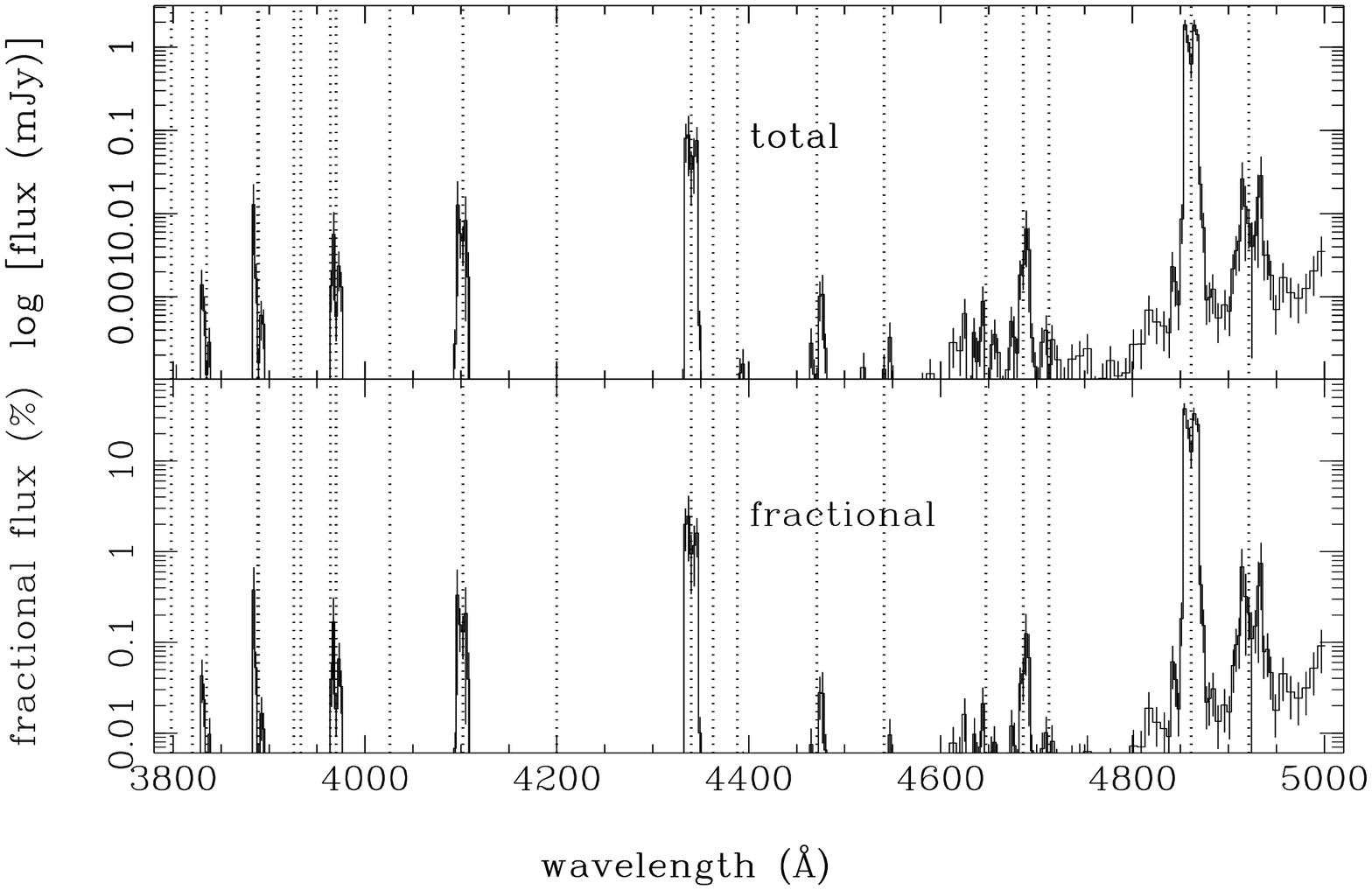}
 \caption{Spectrum of the uneclipsed component. Top: total
     contribution. Bottom: fractional contribution. The notation is
     similar to that of Fig.~\ref{spec_disk}.} \label{neclip}
\end{figure}

Fig.~\ref{neclip} shows the uneclipsed component and its fractional
contribution with respect to the average out-of-eclipse level. The
uneclipsed spectrum is dominated by the Balmer lines, which appear in
emission with an apparent double-peaked profile. H$\beta$ is the most 
prominent line, reaching $\simeq 25\,\%$ of the total emission at its
wavelength. The He\,I $\lambda 4921$ line is also strong in the spectrum,
at a level similar to that of H$\delta$ but with a wider line profile.
In contrast to the observed in the average spectra (in- and 
out-of-eclipse), where the continuum is almost flat, the continuum in the
uneclipsed component rises towards longer wavelengths, reaching the same
intensity level of some lines (e.g., the H$\epsilon$ line). A close
inspection indicate that the uneclipsed Balmer lines have a shallow and
narrow central absorption core (FWHM~$\simeq 200\,km\,s^{-1}$) on top of a
broader emission component (FWHM~$\simeq 800\,km\,s^{-1}$). The absorption
core of the He\,I $\lambda 4921$ line is deeper and wider, with a 
FWHM~$\simeq 600\,km\,s^{-1}$.

It is unlikely that the uneclipsed emission lines are from the secondary
star. Although trailed spectra of Balmer lines show an emission
component with the secondary star motion, the emission comes from the
inward-facing hemisphere of the secondary star --- indicating it is caused
by irradiation effects --- and is apparently restricted to orbital phases
around $\phi\sim 0.5$ --- suggesting that the emitting region is not
visible during eclipse \citep{1995ApJ...448..380M}.  As the inward-facing
hemisphere of the secondary star is not seen around eclipse, it should
not contribute to the uneclipsed light. Furthermore, the width of the
uneclipsed lines is much larger than expected for a rotationally broadened
line from the secondary star in DQ~Her \citep[$V_{\rm rot}\sin i = 115~\pm 
18\,km\,s^{-1}$,][]{1993ApJ...410..357H}.

The uneclipsed emission lines are probably from the nova shell. We note that
nebular emission from the extended nova shell was included in the  slit
together with the light from the binary, and its contribution was not removed
from the spectra \citep{1995ApJ...448..380M}. Accordingly, the  FWHM of the
uneclipsed emission lines is consistent with the expansion velocity of the
nova shell \citep[$v= 370\pm 14\,km\,s^{-1}$, see][]{2007MNRAS.380..175V}.
The narrow absorption component possibly arises in a collimated and optically
thick wind from the accretion disk.  Evidences for the presence of a
collimated disk wind in DQ~Her come from the eclipse behavior of the C\,IV
$\lambda 1550$ resonance line \citep{1998ASPC..137..438E}, the study of
scattered x-ray emission \citep{2003ApJ...594..428M}, and from the ellipsoidal
shape of the nova shell and the ablation of gas clumps around the shell poles
\citep{2007MNRAS.380..175V}.  The secondary star may be responsible for the
rising continuum at the red end of the spectrum, although the short wavelength
range where the continuum appears prevents a meaningful attempt to infer its
spectral type.

\subsection{The temperatures of the disk} \label{temperatures}

\begin{figure}
\includegraphics[bb=1cm -6cm 19cm 20cm,angle=-90,scale=0.44]{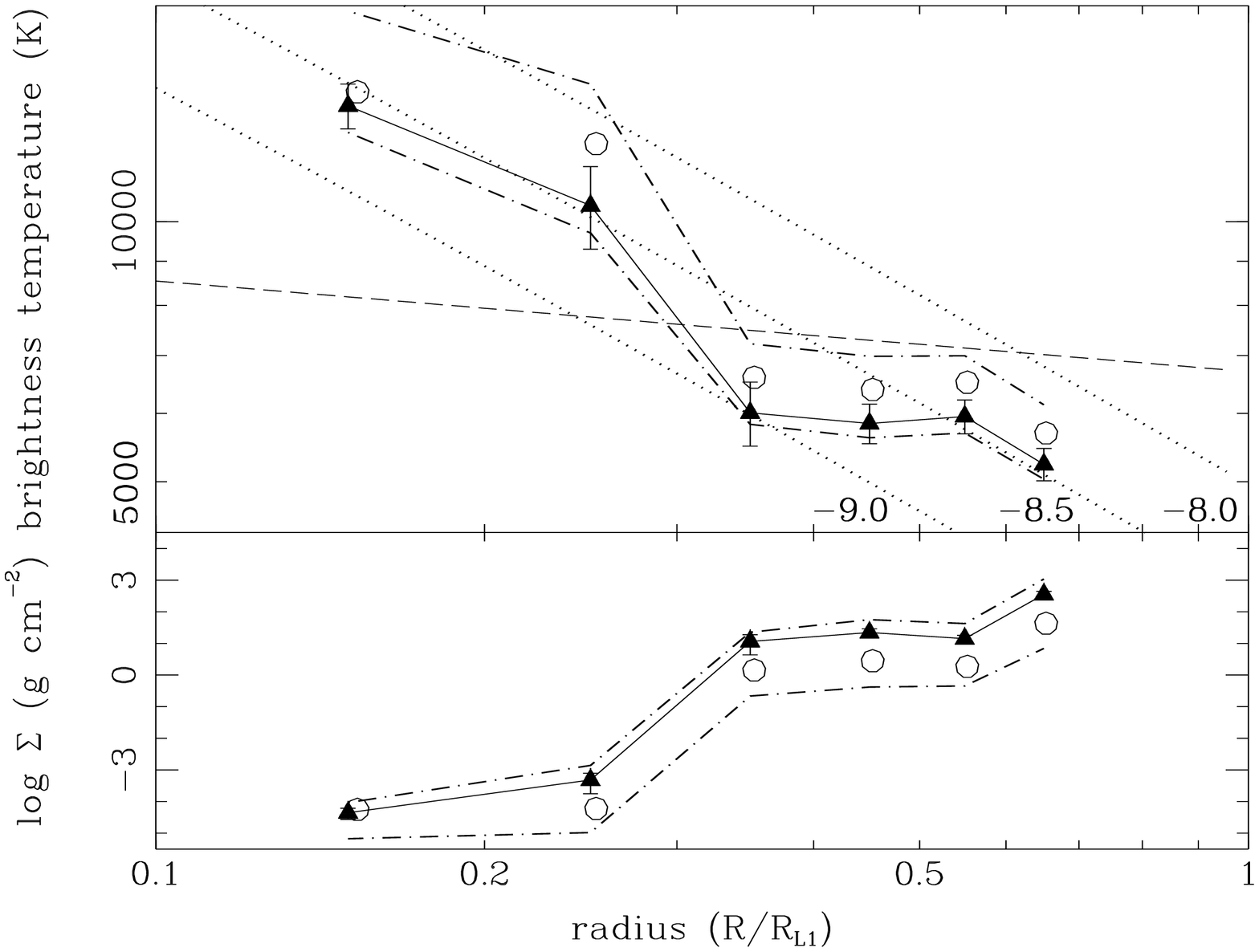}
 \caption{Top: The radial run of the brightness temperature of the 
   best-fit H\,I model to the disk spectra ($d=400\,pc$ and $\beta= 
   2\degr$, solid line with filled triangles) and for $d=525\,pc$ and
   $\beta= 2\degr$ (open circles). The results for two alternative,
   extreme cases of low ($d=525\,pc$ and $\beta=0\degr$) and high 
   ($d=400\,pc$ and $\beta=3.5\degr$) solid angles (see text) are shown
   as the upper and lower dot-dashed curves, respectively. Dotted lines
   show steady-state disk models for mass accretion rates of $\dot{M} =
   10^{-8.0}$, $10^{-8.5}$ and $10^{-9.0}$ M$_\odot{\rm \:yr^{-1}}$.
   The dashed line shows the critical temperature above which the gas
   should remain in a steady, high-\.{M} regime, according to the disk
   instability model. Bottom: the radial run of the surface density. 
   The notation is the same as for the upper panel.} \label{temp}
\end{figure}

Here we fit isothermal, pure hidrogen atmosphere models (H\,I), to the
spatially resolved spectra at each disk annular region (Fig.~\ref{spec_disk})
in order to derive the radial dependence of temperature and column density in
the disk (Fig.~\ref{temp}).  We used the symmetric component of the emission
and we masked the regions of the lines, since the models only account for the
continuum emission produced by bound-free and free-free H\,I transitions. The
fit depends on the solid angle $\theta^2$ assumed for the pixel in the eclipse
maps, which by its turn depends on the distance and the inclination with which
the pixel is seen by an observer at the Earth.  The effective inclination of a
pixel in the far side of the accretion disk is $i_{eff}= i - \beta$, where
$\beta$ is the disk half-opening angle. We obtained fits for distances of
$400\,pc$ \citep{1993ApJ...410..357H} and $525\,pc$
\citep{2007MNRAS.380..175V}, and half-opening angles of $\beta= 0\degr$,
$2\degr$, and $3.5\degr$. 

Fig.~\ref{temp} shows the resulting radial dependence of the temperature and
column density for the case $d=400\,pc$ and $\beta= 2\degr$ (solid line with
filled triangles) and $d=525\,pc$ and $\beta= 2\degr$ (open circles). Also
shown are the results for two extreme cases of low (larger distance and lower
$\beta$) and high (lower distance and larger $\beta$) $\theta^2$. The $\chi^2$
of the fit increases perceptibly for $d= 525\,pc$, because the high
temperatures required to match the average flux level of the spectra can no
longer reproduce the red slope of the observed spectra. Although the different
choices of $\theta^2$ lead to systematic differences in temperature and column
density, these differences are of the order of the uncertainties affecting the
derived quantities and do not alter the inferences that can be drawn from
Fig.~\ref{temp}. Our results (for $d=400\,pc$ and $\alpha= 2\degr$) indicate
that the surface density increases with increasing radius, showing values
of~$\Sigma \sim 4\times10^{-5}\,g\,cm^{-2}$ at $R=0.15 \,R_{L1}$ and $\Sigma
\sim 3\times10^{2}\,g\,cm^{-2}$ at $R=0.65\, R_{L1}$. The steady-state disk
models predict that the surface density is proportional to $\alpha^{-4/5}
R^{-3/4}$, where $\alpha$ is the viscosity parameter. Thus, our values for
$\Sigma$ are consistent with steady-state model only if the viscosity in the
disk decreases with increasing radius in order to compensate the factor
$\propto R^{-3/4}$. The brightness temperature ranges between
$T\simeq13500\,K$ at $R=0.15\,R_{L1}$ and $T\simeq5000\,K$ at
$R=0.65\,R_{L1}$, with a radial dependence reasonably well described by a
steady-state optically thick disk model with mass accretion rate of
$\dot{M}=(2.7\pm1.0)\times10^{-9} \,M_{\odot}\,yr^{-1}$. The inferred mass
accretion rate is in reasonable agreement with the values derived by
\cite[][$\dot{M}= 10^{-9}\,M_{\odot}\,yr^{-1}$]{1978ApJ...226..963C} and
\cite[][$\dot{M}=3.4\times10^{-9}\,M_{\odot}\,yr^{-1}$]{1995ApJ...454..447Z}. 

Temperatures in excess of $8000\,K$ are required in order to produce the
strong lines with shallow Balmer decrement observed in the intermediate
and outer disk regions. This is in contrast with the low temperatures
($T \lesssim 5500\,K$) inferred from the slope of the red continuum
emission, suggesting the existence of a vertical temperature gradient
in the DQ~Her disk gas, with a hot, optically thin chromospheric layer 
(responsible for the emission lines) on top a cool, opaque and dense disk
photosphere (responsible for the red continuum). The chromosphere is
likely heated by irradiation of x-ray photons from the magnetic poles
and accretion curtains, leading to high-excitation lines. This scenario
is beyond LTE simple models.

\cite{1981AcA....31..395S} remarks that in DQ Her the continuum emission seems
to be a function of the accretion disk regime while the line emission appears
to be correlated with the luminosity or surface temperature of the primary
star. In this context the lines are formed by reprocessing of the energy
radiated from the central source, probably the accreting material falling over
the magnetic poles of the white dwarf. This central emission in DQ Her can be
detected in soft x-ray emission \citep{2003ApJ...594..428M}.

We note that the temperatures in the outer disk regions ($R \gtrsim
0.3\,R_{L1}$) are systematically below the critical temperature for disk
instabilities to set in \citep{1995cvs..book.....W}, suggesting that the DQ
Her disk could show dwarf nova-type outbursts. This holds even if one assumes
the higher mass accretion rate of \cite{1995ApJ...454..447Z} (illustrated by the
steady-state disk model with $\dot{M} = 10^{-8.5}$ in Fig.~\ref{temp}).
There is no record of dwarf nova outbursts in this binary. 

We end this section raising a puzzlingly different scenario from the radial
temperature distribution of Fig.~\ref{temp}. Given the evidences presented
here of magnetically-controlled accretion for $R\lesssim 0.3 \,R_{L1}$, it is
reasonable to assume that a {\it bona fide} accretion disk exists only for
distances outside this range, and that the brake in slope of the temperature
and column density distributions at $R \sim 0.3 \,R_{L1}$ reflects the change
from the azimuthal flow of gas in the outer accretion disk to a
magnetically-controlled radial flow of gas over the accretion curtains in the
inner regions.  In this case, the outer accretion disk shows a flat
temperature distribution resembling those found in quiescent dwarf novae
\citep[e.g.,][]{1986MNRAS.219..629W, 1992ApJ...385..294W}. The inferred low
temperatures ($T\sim 6000\,K$) and high surface densities ($\Sigma \sim 10^4\,
g\,cm^{-2}$) suggest that the disk viscosity is low.  This is in marked
contrast to the common view that the old nova DQ~Her is a nova-like variable
with a hot, viscous accretion disk.

\section{Discussion}\label{discussion}

\subsection{An energy dependency of the emission-line radial 
intensity distribution slope}
\label{energy-slope}

\begin{figure}
\includegraphics[bb=1.5cm -7cm 20cm 18cm,angle=-90,scale=0.40]{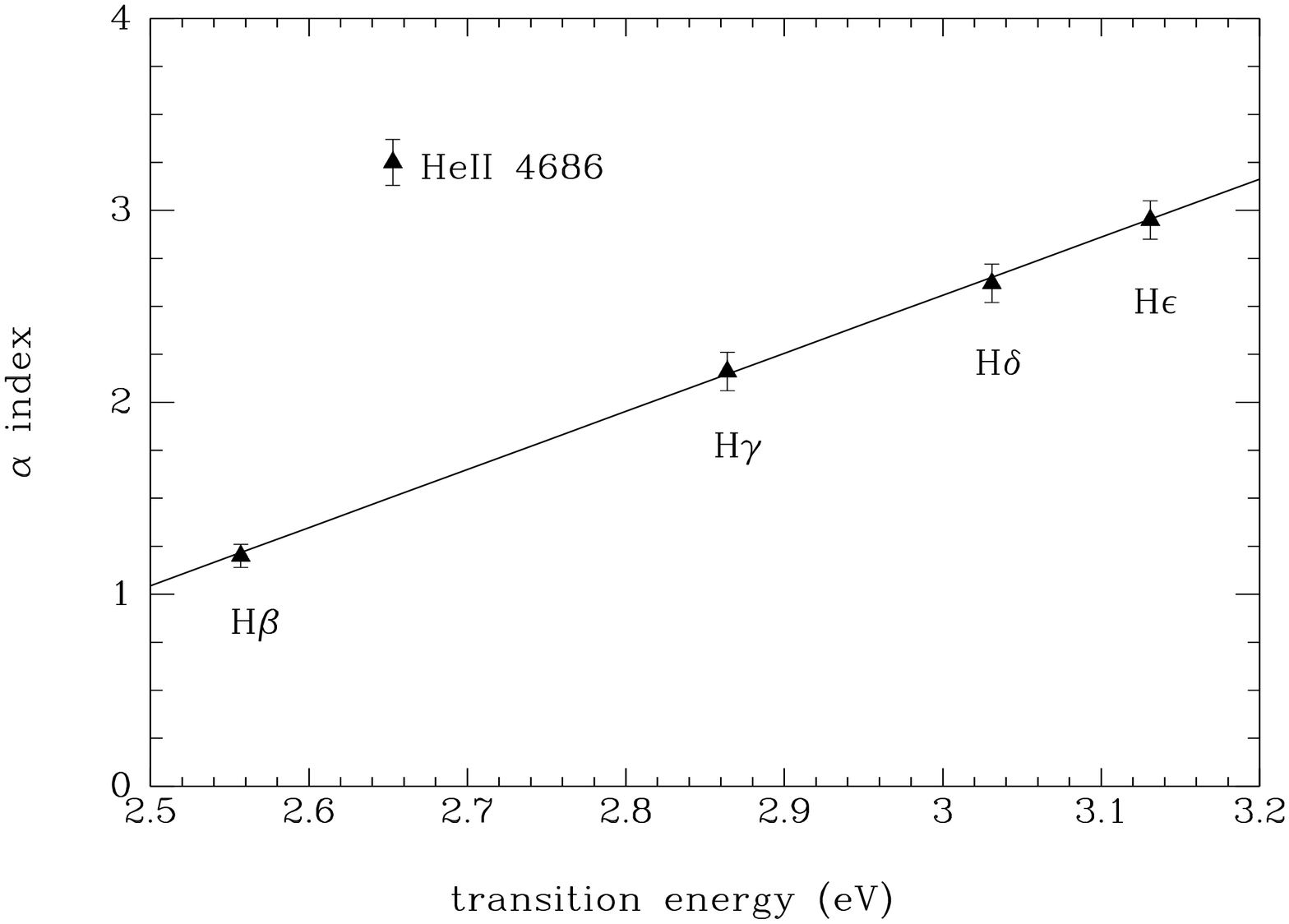}
 \caption{Distribution of the index $\alpha$ in function of the transition
   energy for the He\,II 4686 and Balmer series lines. The solid line marks a
   linear fit for the Balmer lines distribution (see
   Section~\ref{energy-slope}).}
\label{alphas}
\end{figure}

In Section~\ref{emission} we found that the slope of the radial line intensity
distributions correlates with the transition energy of the Balmer lines.  In
order to extend our analysis we extracted (using the same procedures described
in Section~\ref{emission}) the radial intensity distribution for H$\epsilon$,
the next line in the Balmer series. We then fitted a power-law $I \propto
r^{-\alpha}$ \citep[e.g.,][]{1981AcA....31..395S, 1986MNRAS.218..761H} for
each radial distribution in order to find the best-fit $\alpha$ index for each
line. The innermost disk regions ($R < 0.2~R_{L1}$) are excluded from the fit
since the continuum reaches the emission level of the lines in the regions
inside this radius. The results are listed in Table~1 and plotted in
Fig.~\ref{alphas}.

\begin{table}
\begin{center}
\caption{Index $\alpha$ for the Balmer series and He\,II lines.}
\vspace{12 pt}
\begin{tabular}{l c c c c}  
\hline
\hline
  Line        & E[eV] & $\alpha$ ($I \propto r^{-\alpha}$) \\ 
\hline
  H$\beta$    & 12.76 &  $1.20 \pm 0.06$ \\
  H$\gamma$   & 13.06 &  $2.16 \pm 0.11$ \\
  H$\delta$   & 13.23 &  $2.62 \pm 0.10$ \\
  H$\epsilon$ & 13.33 &  $2.95 \pm 0.12$ \\
  He\,II 4686 & 75.50 &  $3.15 \pm 0.14$ \\
\hline
\end{tabular}
\end{center}
\end{table}

The $\alpha$ index increases with the transition energy for the Balmer
lines, i.e., the emission at lines with larger transition energy (e.g.,
H$\epsilon$) are more concentrated at the disk center than lines of
lower energy (e.g., H$\beta$). Moreover, our results reveal that there is
a strong linear correlation between $\alpha$ and the transition energy
for the Balmer lines (see Fig.~\ref{alphas}). This result is in contrast
with the empirical $I \propto r^{-1.5}$ law \citep{1990ApJ...364..637M} usually
evoked in works about radial distribution of emission lines in accretion
disks. The He\,II 4686 line radial distribution falls much more rapidly
with radius than the Balmer lines, with an $\alpha = 3.15$, and does not
fit the linear relation of the Balmer lines. This is not surprising, as
significantly different physical conditions are required to produce the
H\,I and He\,II lines.

\cite{1969AcA....19..155S} and \cite{1972ApJ...171..549H} showed that the
double-peaked profile of emission lines in accretion disks can be accounted
for by making two basics assumption: ($i$) the geometry is flat, i.e., the
disk emission is restricted to the orbital plane, and ($ii$) the disk is in
Keplerian motion. \cite{1981AcA....31..395S} showed that the line profile is
sensitive to the line radial intensity distribution and that the $\alpha$
value could be recovered by fitting a model to the line profile.  From
observational evidences
\citep[e.g.,][]{1979ctvs.conf...65F,1980ApJ...238..955Y} he found that the
emission lines profiles could be represented by power-laws $I \propto
r^{-\alpha}$, where $\alpha$ is expected to be in the range $1-2$.

\cite{1981AcA....31..395S} gives a list of results where $\alpha$ could be
estimated for Helium and Balmer series lines (Z~Cha,
\citealt{1979ctvs.conf...65F}; SS~Cyg,
\citealt{1980ApJ...241..269C,1980ApJ...240..597S}; HT~Cas,
\citealt{1981ApJ...245.1035Y}; DQ~Her,
\citealt{1979ApJ...232..500H,1980AZh....57..749D}). In all cases the value
obtained for $\alpha$ is in the range $1.7-2.2$ suggesting that there are no
systematic differences between objects of different classes of cataclysmic
variables. Similar values of $\alpha$ are obtained by fitting the line profile
of the dwarf novae U~Gem or Z~Cha and the intermediate polar
DQ~Her. \cite{1979ApJ...232..500H} found that in DQ~Her the H$\beta$ line has
$\alpha=2.2$ while He\,II 4686 has $\alpha=1.9$.  These values are
inconsistent with our results. The discrepancy may be explained by the low
spectral resolution of the data of \cite{1979ApJ...232..500H}, which obtained
$\alpha$ from line profiles of low resolution (dispersion of 48 \AA\,
mm$^{-1}$) while we extracted $\alpha$ from spatially resolved spectra
obtained from data of significantly higher spectral resolution (2
\AA\,pixel$^{-1}$).

\cite{1987MNRAS.228..779M} analysed the Balmer lines in the dwarf nova Z
Cha. In his study the differences between the line profiles in the average
spectrum are evident. The line wings become wider from H$\alpha$ to H$\delta$
indicating that the radial line distributions become more concentrated for
increasing transition energy. In quiescence, Z Cha has a relatively small
inferred mass transfer rate of $\dot{M}=10^{-12}\,M_{\odot}\,
yr^{-1}$. Simulations shows that the LTE model is not valid in this case,
producing a poor fit to the Balmer lines \citep{1987MNRAS.228..779M}.

\cite{1988MNRAS.231.1117M} found $\alpha=1.8$ for the Balmer lines in the
dwarf nova IP~Peg in quiescence. Similarly, \cite{1990ApJ...349..593M} showed
that in IP~Peg the radial distribution of the Balmer lines have similar slopes
when the object is in quiescence but that the radial distributions are
different for distinct lines when the object is in outburst. The H$\delta$
line distribution is steeper than H$\gamma$, that is steeper than H$\beta$,
i.e., the $\alpha$ index of the radial profile increases with the transition
energy. A similar result was found by \cite{1990ApJ...364..637M} for the dwarf
nova U~Gem, with H$\gamma$ and H$\beta$ showing, respectively, $\alpha=1.66$ e
$1.43$. In contrast, for the white dwarf V2051~Oph, \cite{2006AJ....131.2185S}
found $\alpha=1.55$ for H$\delta$ and $\alpha=1.78$ for H$\gamma$; $\alpha$
decreases with the transition energy in this dwarf nova. During their
observations V2051~Oph was in an unusual low brightness state (low mass
transfer rate). Thus, there are observational evidences that the slope of the
radial line distribution is different for different emission lines.

In order to understand the physics behind the line formation in accretion
disks it is necessary to combine new and improved theory with 
spatially-resolved observational studies. Spatially resolved studies from
Doppler tomography and eclipse mapping techniques can provide high
resolution information about the line distribution in the accretion disk.
More sophisticated spectral models for accretion disk need to include
features such as the vertical temperature structure and external sources
of heat.

\subsection{Are we seeing the accretion curtains?}
\label{curtains}

Here we discuss possible interpretations for the asymmetries seen in the
low-velocity eclipse maps of the H$\beta$ and He\,II $\lambda 4686$ lines.
These asymmetries cannot be explained in terms of gas stream overflow
\citep[e.g.,][see their Fig.~5]{2004AJ....128..411B} as they are  physically
unrelated to the gas stream trajectory, nor can they be interpreted in terms
of spiral shock waves \citep[e.g.,][see their Fig.~2]{2000MNRAS.314..727B} as
the morphology, orientation and location are inconsistent with those expected
for two-armed  tidally-induced spiral structures. Moreover, it is also not
possible to explain the observed structures in terms of a front-back asymmetry
caused by a flared disk \citep[e.g.,][see their Figs.~3 and
8]{1998MNRAS.298.1079B} because this would affect both line and continuum
maps, whereas in DQ~Her the asymmetries are only seen at specific
velocity-resolved eclipse maps of particular emission lines. Additionally, the
disk opening angle is not large enough to result in measurable obscuration or
limb-darkening effects (see Section~\ref{mem}).  The asymmetries observed in
the H$\beta$ and He\,II $\lambda 4686$ line maps seem to be of a different
nature, related to something not previously seen in eclipse mapping
experiments.

\begin{figure}
\includegraphics[bb=1cm -15cm 20cm 18cm,angle=-90,scale=.3]{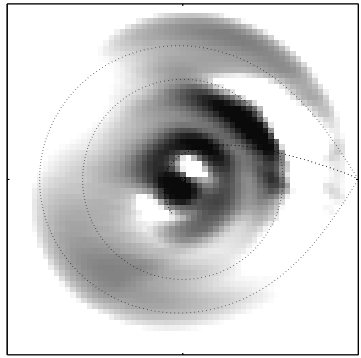}
\caption{Combined net emission He\,II plus H$\beta$ low-velocity eclipse
map. The notation is similar to that of Fig.~\ref{hbet_curves}.}
\label{disk_soma}
\end{figure}

DQ~Her is the first intermediate polar being eclipse mapped. It is tempting to
interpret the structures seen in the low-velocity He\,II and H{$\beta$} maps
in terms of the projection onto the orbital plane of accretion
columns/curtains formed over the magnetic poles of the white dwarf by the
channeled material \citep[e.g.,][]{1996PASA...13...87F}\footnote {The eclipse
mapping procedure has the approximate effect  of projecting any emission from
above the orbital plane back along the  inclined line of sight to the point
where it pierces the disk plane.  Thus, emission arising at height $Z$ above
the disk is projected a  distance $\rm Z \, {\it tan}\, \rm i$ towards the
back of the disk and is distorted by the intrinsic azimuthal smearing effect
of the eclipse mapping method \citep{1995ApJ...448..395B}. With our choice of
azimuthal blur width $\Delta\phi$ (Section~\ref{mem}), any point source in the
eclipse maps will appear smeared in azimuth by $\simeq 20\degr$.}. In this
scenario, due to different opacities for incoming irradiating photons, and
energies needed to ionize atoms and to excite electrons to the upper energy
levels of each transition, the regions of the accretion curtains close to the
magnetic poles are prone to reprocess the strong irradiation by hard x-rays
photons whereas regions farther away receive less energetic photons. Hence,
high-excitation lines such as He\,II $\lambda 4686$ trace the emission in the
accretion curtains regions closer to the magnetic poles, whereas  the
low-excitation Balmer lines trace emission from regions farther away from the
x-ray emitting magnetic poles. Lines of different excitation frame different
parts of the accretion flow and, individually, yield only a partial view of
the emitting accretion curtains. According to this, a better view of the
accretion curtains may be obtained by combining the complementary
high-excitation He\,II map with the low-excitation H$\beta$ map. An eclipse
map obtained by combining the net line emission He\,II $\lambda 4686$ and
H$\beta$ low-velocity eclipse maps is shown in  Fig.~\ref{disk_soma}. A
remarkable structure resembling a twisted dipole emitting pattern can be seen
near disk center. Although this figure encourages one to further pursue this
line of reasoning, it poses an immediate problem.

Each eclipse map yields a snapshot of the disk brightness distribution
averaged over the period covered by the 4 eclipses. Therefore, because
it comprises a timescale much longer than the 71\,$s$ spin period of the 
white dwarf, changes in brightness distribution caused by the varying
aspect of the accretion curtains with the white dwarf rotation should be
completely smeared out in the eclipse map --- just as in a long 
exposure picture of a moving target. How could one then obtain a
coherent image of fast rotating accretion curtains in an eclipse map?
The answer can be built in terms of the selection in velocity for the
accreting material performed by the application of spectral mapping
techniques.

The velocity field for gas in the accretion curtains is a sum of its inflowing
component (along the magnetic field lines) and a rotational  component (due to
the white dwarf spin period). As seen in
Sections~\ref{spectra}~and~\ref{emission}, the velocity field is dominated by
the rotational component, which is significantly larger than the  inflowing
component. At most rotation phases (i.e., curtain orientations), different
parts of the curtains show distinct projected velocities along the line of
sight and their emission will split onto several different velocity bins
across the emission line profile.  Conversely, a specific velocity-resolved
eclipse map will contain a mixture of different parts of the accretion
curtains at different rotation phases. This slicing and mixing effect prevents
one from obtaining a coherent image of the accretion curtains at all
velocity-resolved eclipse maps, with the fortunate exception of the
low-velocity bin. 

When the curtains are aligned with the binary axis --- and only at this
orientation --- the rotational component will be suppressed (because it
becomes orthogonal to the line of sight), and the curtains will show only
the small inflowing velocity component (see Section~\ref{spectra}),
contributing to the same particular low-velocity bin. 
We note that two effects contribute to further reduce the observed
projected velocity: (i) the inflow velocity must decrease after the
accretion shock until it reaches zero at the white dwarf surface, and
(ii) for accretion curtains twisted off the orbital plane, regions
close to the magnetic poles present the lowest projected velocity
along the line of sight.
Therefore, looking at a low velocity bin is equivalent of framing the
accretion curtains always at the same two spin phases (where the curtains
are oriented at an azimuth roughly aligned with the binary main axis),
allowing one to obtain a coherent image of them.

A further question is why the accretion curtains lag the binary main
axis by $\simeq 30\degr$. \cite{1999ApJ...519..324H} applied Doppler tomography
techniques to spin-pulse spectroscopy of intermediate polars to find
that the optical pulsation arises from azimuthally-extended accretion
curtains, located typically a few white dwarf radii from disk center. 
His tomogram of the slow rotator PQ~Gem shows a twist in the accretion
curtains, with material feeding onto magnetic field lines $\sim 40\degr$
ahead of the location at which it hits the white dwarf. This is the
opposite case of what is seen in DQ~Her and the difference may be in the
spin period. For comparable magnetic field strengths and mass accretion
rates a slow rotator (the case of PQ~Gem, $P_{spin}=834\,s$) picks
material at a magnetospheric radius with higher (Keplerian) velocity
than the accretion curtains. Thus, the accretion curtains tend to step
ahead of the magnetic pole. On the other hand, in a fast rotator (the
case of DQ Her, $P_{spin}=71\,s$) the field lines at the magnetosphere
are rotating faster than the disk gas being channelled onto the field
lines (see Section~\ref{emission}). In this case the accretion curtains tends to
lag the magnetic pole.

\section{Summary}\label{summ}

We applied eclipse mapping techniques to spectroscopy covering 4 eclipses
of DQ~Her in order to isolate the emission from different parts of its
accretion flow as well as from outside the orbital plane. The main 
results of this study can be summarized as follows:

\begin{enumerate}

\item Velocity-resolved eclipse maps of the H$\beta$ and He\,II 
$\lambda 4686$ lines indicate the presence of an azimuthally- and 
vertically-extended bright spot at disk rim (at R$_{BS}= 0.57\pm 0.03\,
R_{L1}$), which is an important source of reprocessing of x-rays from
the magnetic poles. This is in line with the conclusion that the bright
spot is the main site of the $71\,s$ optical pulsations in DQ~Her
\citep{2009ApJ...693L..16S}.

\item In the inner regions the disk spectrum is flat with no Balmer 
or Helium lines, suggesting that the emission arises from 
{\it bremsstrahlung} radiation by an optically thin gas. In the 
intermediate and outer disk regions the spectrum shows double-peaked
emission lines typical of a rotating disk gas. The slope of the 
continuum becomes progressively redder with increasing radius,
indicating the existence of a radial temperature gradient.

\item We fit LTE H\,I emission models to the flux and slope of the
continuum in order to infer the temperature and the surface density of
the emitting gas.
The surface density increases by six orders of magnitude from the inner
to the outer disk regions. The temperatures are in the range $T\simeq
13500 - 5000~K$ (for $R= [0.15 - 0.65]\,R_{L1}$) and can be reasonably
well described by a steady-state disk with a mass accretion rate of
$\dot{M}=(2.7\pm 1.0) \times 10^{-9}\,M_{\odot}\,yr^{-1}$. 
The strong emission lines with shallow Balmer decrement observed in the
outer disk regions are in contrast to the low temperatures inferred
from the slope of the continuum, suggesting an outer disk structure
with a hot and optically thin chromosphere (responsible for the emission
lines) on top of a cool, opaque and dense disk photosphere (responsible
for the continuum emission).

\item The narrow and redshifted Ca\,II $\lambda 3934$ absorption line
in the total light spectra and the inverse P-Cygni profiles of the
Balmer and He\,II $\lambda 4686$ emission lines in spectra of the
asymmetric component of the eclipse maps indicate radial inflow of 
gas in the inner disk regions ($R< 0.3\,R_{L1}$), and are
best explained in terms of magnetically-controlled accretion inwards
of the white dwarf magnetosphere in DQ~Her. We infer projected radial
inflow velocities of $(200-500)\,km\,s^{-1}$, significantly lower than
both the rotational and the free-fall velocities for the corresponding
range of radii.

\item The combined net emission He\,II plus H$\beta$ low-velocity eclipse
map shows a twisted dipole emitting pattern near disk center. This is
interpreted as being the projection of accretion curtains onto the 
orbital plane at two specific spin phases, as a consequence of the
selection in velocity provided by the spectral eclipse mapping. This
is in line with and strengthen the above inference.

\item The spectrum of the uneclipsed light is dominated by Balmer and
He\,I lines in emission with narrow absorption cores. The line emission 
is probably from the extended nova shell, whereas the narrow absorption
likely arises in a collimated and optically thick wind from the accretion
disk.

\item A comparison of the line radial distribution for the Balmer lines
  reveals a linear correlation between the slope of the distribution and 
 the transition energy.

\end{enumerate}

\acknowledgments

The white dwarf atmosphere models used in the work were kindly provided by
Dr. Detlev Koester. We thank Jo\~ao Steiner for an enlightening discussion
which led to the novel application of eclipse mapping techniques to map
the $71\,s$ optical pulsations in DQ~Her \citep{2009ApJ...693L..16S}.
This work was partially supported by CNPq/Brazil through the research grant
62.0053/01-1-- PADCT III/Milenio. RS acknowledges financial support from CONICYT
through GEMINI Project Nr. 32080016, BASAL PFB-06, FONDAP Center for
Astrophysics Nr. 15010003 and CNPq/Brazil. RB acknowledges financial support
from CNPq/Brazil through grants n. 300.354/96-7 and 301.442/2004-5.

\appendix

\section{Reconstruction of FWHM distributions of 
  emission lines}\label{appendixA}

Here we address the reliability of spectral mapping techniques to
reconstruct FWHM distributions of emission lines from spatially
resolved disk spectra.  Our aim is to check whether the sub-Keplerian
velocities found for the Balmer and He\,II 4686 lines in DQ~Her
(Section~\ref{emission}) are intrinsic to the variable or an artifact of the
mapping method.

\begin{figure}
\includegraphics[bb=-6cm 0.5cm 14cm 24cm,scale=0.57]{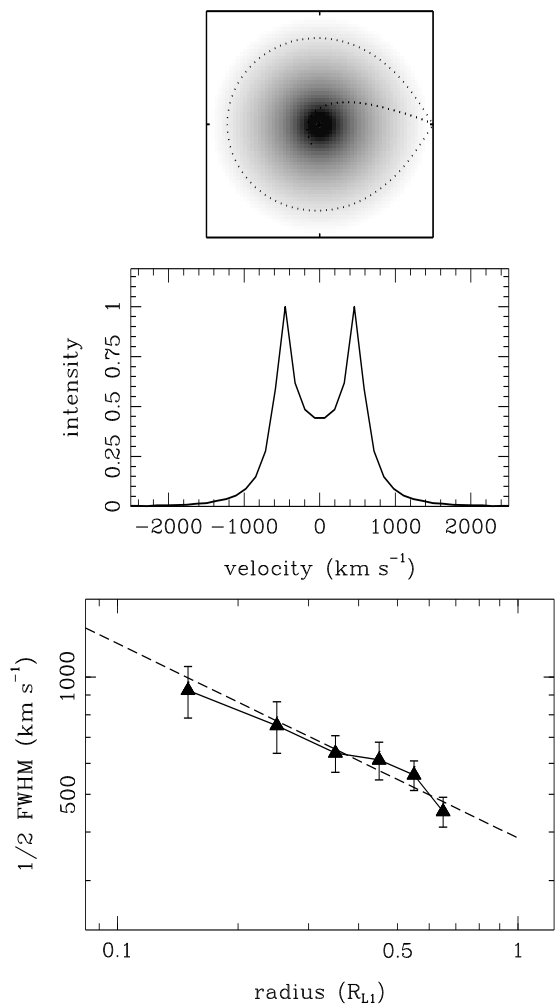}
\caption{Top: Greyscale plot of the brightness distribution of the
steady-state disk used in the simulation. Middle: the emission line
profile derived from the brightness distribution of the top panel
assuming a Keplerian velocity field according to the parameters of
DQ Her. Bottom: The FWHM of the emission line as a function of radius
derived from the spectral mapping of 23 light curves along the emission
line profile of the middle panel. A dotted line shows the $v \propto
r^{-1/2}$ law expected for Keplerian orbits around a white dwarf of mass
$M_{1}=0.6\,M_{\odot}$.}
\label{kepler}
\end{figure}

First, we created a synthetic eclipse map of a steady-state opaque accretion
disk according to the set of parameters of DQ~Her: white dwarf mass of
$M_{1}=0.6\,M_{\odot}$ \citep{1995ApJ...454..447Z}, white dwarf radius of
$R_{WD}=0.0121\,R_{\odot}$ \citep{1995ApJ...448..380M}, Roche lobe size of
$R_{L1}=0.766\,R_{\odot}$ \citep{1980ApJ...241..247P} and an accretion rate of
$\dot{M}= 2.7 \times10^{-9}\,M_{\odot}\,yr^{-1}$ (Section~\ref{temperatures}).
We adopted the geometry of DQ~Her \citep[$i = 86.5\degr$ and $q =
0.66$,][]{1993ApJ...410..357H} to simulate the eclipse of this brightness
distribution and to generate velocity-resolved synthetic light curves for the
same set of bins used in the spectral mapping of DQ~Her
(Section~\ref{lcurves}). In order to construct the light curve of a given
velocity bin, we assumed a Keplerian velocity field for the synthetic eclipse
map and computed, at each orbital phase, the flux contribution only for those
pixels the projected (Doppler) velocity of which falls inside the velocity bin
under consideration.  The intensity distribution of the synthetic map is shown
in the upper panel and the corresponding double-peaked emission line profile
is depicted in the middle panel of Fig.~\ref{kepler}.

Gaussian noise was added to the synthetic light curves to simulate
the signal-to-noise ratio of the DQ Her data and eclipse mapping
techniques were applied to the resulting light curves to generate a
set of velocity-resolved synthetic eclipse maps (see Section~\ref{mem}).
We combined the velocity-resolved maps to derive spatially-resolved
spectra for the same set of radial annuli used in the analysis of the
DQ~Her data, and we extracted the FWHM distribution of the emission
line as a function of radius following the same procedure used in
Section~\ref{emission}. The lower panel of Fig.~\ref{kepler} compares the
derived FWHM distribution with the $v \propto r^{-1/2}$ law used to
construct the original eclipse map. Uncertainties were derived via
Monte Carlo simulations with the synthetic light curves.

The derived FWHM distribution is in good agreement with the Keplerian
expectation (at the 1$-\sigma$ confidence level) at all radii. These
simulations show that the FWHM distribution of emission lines can be
reliably recovered from spatially-resolved spectra of accretion disks
obtained with eclipse mapping techniques.
This exercise indicates that the sub-Keplerian velocity distributions
observed in the DQ Her lines are intrinsic to the variable and not
an artifact of the mapping procedure.


\begin{thebibliography}{}

\bibitem[Africano \& Olson(1981)]{1981PASP...93..130A} Africano, J.~L., \&
  Olson, E.~C.\ 1981, \pasp, 93, 130 
\bibitem[Baptista(2001)]{2001LNP...573..307B} Baptista, R.\ 2001, 
Astrotomography, Indirect Imaging Methods in Observational Astronomy, 573, 
307
\bibitem[Baptista \& Catal{\'a}n(2001)]{2001MNRAS.324..599B} Baptista, R., \&
  Catal{\'a}n, M.~S.\ 2001, \mnras, 324, 599 
\bibitem[Baptista et al.(2000)]{2000MNRAS.314..727B} Baptista, R., 
Harlaftis, E.~T., \& Steeghs, D.\ 2000, \mnras, 314, 727 
\bibitem[Baptista et al.(1995)]{1995ApJ...448..395B} Baptista, R., Horne, 
K., Hilditch, R.~W., Mason, K.~O., \& Drew, J.~E.\ 1995, \apj, 448, 395 
\bibitem[Baptista et al.(1998)]{1998MNRAS.298.1079B} Baptista, R., Horne, 
K., Wade, R.~A., Hubeny, I., Long, K.~S., 
\& Rutten, R.~G.~M.\ 1998, \mnras, 298, 1079 
\bibitem[Baptista \& Steiner(1993)]{1993AA...277..331B} Baptista, R., \&
  Steiner, J.~E.\ 1993, \aap, 277, 331 
\bibitem[Baptista et al.(1996)]{1996MNRAS.282...99B} Baptista, R., Steiner, 
J.~E., \& Horne, K.\ 1996, \mnras, 282, 99 
\bibitem[Baptista \& Bortoletto(2004)]{2004AJ....128..411B} Baptista, R., \&
  Bortoletto, A.\ 2004, \aj, 128, 411 
\bibitem[Chanan et al.(1978)]{1978ApJ...226..963C} Chanan, G.~A., Nelson, 
J.~E., \& Margon, B.\ 1978, \apj, 226, 963 
\bibitem[Cowley et al.(1980)]{1980ApJ...241..269C} Cowley, A.~P., Crampton, 
D., \& Hutchings, J.~B.\ 1980, \apj, 241, 269 
\bibitem[Dmitrienko 
\& Cherepashchuk(1980)]{1980AZh....57..749D} Dmitrienko, E.~S., \&
Cherepashchuk, A.~M.\ 1980, \azh, 57, 749 
\bibitem[Eracleous et al.(1998)]{1998ASPC..137..438E} Eracleous, M., Livio, 
M., Williams, R.~E., Horne, K., Patterson, J., Martell, P., 
\& Korista, K.~T.\ 1998, Wild Stars in the Old West, 137, 438 
\bibitem[Fabian et al.(1979)]{1979ctvs.conf...65F} Fabian, A.~C., Pringle, 
J.~E., Whelan, J.~A.~J., \& Bailey, J.~A.\ 1979, IAU Colloq.~46: Changing
Trends in Variable Star Research, 65 
\bibitem[Ferrario(1996)]{1996PASA...13...87F} Ferrario, L.\ 1996, 
Publications of the Astronomical Society of Australia, 13, 87 
\bibitem[Hellier(1999)]{1999ApJ...519..324H} Hellier, C.\ 1999, \apj, 519, 
324 
\bibitem[Hellier(2001)]{2001cvs..book.....H} Hellier, C.\ 2001, Cataclysmic 
Variable Stars, Springer, 2001,  
\bibitem[Horne(1985)]{1985MNRAS.213..129H} Horne, K.\ 1985, \mnras, 213, 
129 
\bibitem[Horne \& Marsh(1986)]{1986MNRAS.218..761H} Horne, K., \& Marsh,
  T.~R.\ 1986, \mnras, 218, 761 
\bibitem[Horne et al.(1993)]{1993ApJ...410..357H} Horne, K., Welsh, W.~F., 
\& Wade, R.~A.\ 1993, \apj, 410, 357 
\bibitem[Huang(1972)]{1972ApJ...171..549H} Huang, S.-S.\ 1972, \apj, 171, 
549 
\bibitem[Hutchings et al.(1979)]{1979ApJ...232..500H} Hutchings, J.~B., 
Cowley, A.~P., \& Crampton, D.\ 1979, \apj, 232, 500 
\bibitem[Kraft(1959)]{1959ApJ...130..110K} Kraft, R.~P.\ 1959, \apj, 130, 
110 
\bibitem[Marsh(1987)]{1987MNRAS.228..779M} Marsh, T.~R.\ 1987, \mnras, 228, 
779 
\bibitem[Marsh(1988)]{1988MNRAS.231.1117M} Marsh, T.~R.\ 1988, \mnras, 231, 
1117 
\bibitem[Marsh \& Horne(1988)]{1988MNRAS.235..269M} Marsh, T.~R., \& Horne,
  K.\ 1988, \mnras, 235, 269 
\bibitem[Marsh \& Horne(1990)]{1990ApJ...349..593M} Marsh, T.~R., \& Horne,
  K.\ 1990, \apj, 349, 593 
\bibitem[Marsh et al.(1990)]{1990ApJ...364..637M} Marsh, T.~R., Horne, K., 
Schlegel, E.~M., Honeycutt, R.~K., \& Kaitchuck, R.~H.\ 1990, \apj, 364, 637 
\bibitem[Martell et al.(1995)]{1995ApJ...448..380M} Martell, P.~J., Horne, 
K., Price, C.~M., \& Gomer, R.~H.\ 1995, \apj, 448, 380 
\bibitem[Mukai et al.(2003)]{2003ApJ...594..428M} Mukai, K., Still, M., 
\& Ringwald, F.~A.\ 2003, \apj, 594, 428 
\bibitem[O'Donoghue(1985)]{1985...conf...98} O'Donoghue, D., 1985, in
  Proc. Ninth American Workshop on Cataclymic Variables, ed.\ P.\ Szkody
  (Seattle: Univ.\ Washington), p.\,98
\bibitem[Patterson(1994)]{1994PASP..106..209P} Patterson, J.\ 1994, \pasp, 
106, 209 
\bibitem[Patterson et al.(1978)]{1978ApJ...224..570P} Patterson, J., 
Robinson, E.~L., \& Nather, R.~E.\ 1978, \apj, 224, 570 
\bibitem[Petterson(1980)]{1980ApJ...241..247P} Petterson, J.~A.\ 1980, 
\apj, 241, 247 
\bibitem[Rutten et al.(1992)]{1992AA...265..159R} Rutten, R.~G.~M., Kuulkers,
  E., Vogt, N., \& van Paradijs, J.\ 1992, \aap, 265, 159 
\bibitem[Saito \& Baptista(2006)]{2006AJ....131.2185S} Saito, R.~K., \&
  Baptista, R.\ 2006, \aj, 131, 2185 
\bibitem[Saito \& Baptista(2009)]{2009ApJ...693L..16S} Saito, R.~K., \&
  Baptista, R.\ 2009, \apjl, 693, L16 
\bibitem[Smak(1969)]{1969AcA....19..155S} Smak, J.\ 1969, Acta Astronomica, 
19, 155 
\bibitem[Smak(1981)]{1981AcA....31..395S} Smak, J.\ 1981, Acta Astronomica, 
31, 395 
\bibitem[Stover et al.(1980)]{1980ApJ...240..597S} Stover, R.~J., Robinson, 
E.~L., Nather, R.~E., \& Montemayor, T.~J.\ 1980, \apj, 240, 597 
\bibitem[Vaytet et al.(2007)]{2007MNRAS.380..175V} Vaytet, N.~M.~H., 
O'Brien, T.~J., \& Rushton, A.~P.\ 2007, \mnras, 380, 175 
\bibitem[Walker(1956)]{1956ApJ...123...68W} Walker, M.~F.\ 1956, \apj, 123, 
68 
\bibitem[Warner(1986)]{1986ApSS.118..271W} Warner, B.\ 1986, \apss, 118, 271 
\bibitem[Warner(1995)]{1995cvs..book.....W} Warner, B.\ 1995, Cambridge 
Astrophysics Series, Cambridge, New York: Cambridge University Press, 
|c1995, 
\bibitem[Wickramasinghe(1988)]{1988prco.book..199W} Wickramasinghe, D.~T.\ 
1988, Polarized Radiation of Circumstellar Origin, 199 
\bibitem[Wood et al.(1986)]{1986MNRAS.219..629W} Wood, J., Horne, K., 
Berriman, G., Wade, R., O'Donoghue, D., \& Warner, B.\ 1986, \mnras, 219, 629 
\bibitem[Wood et al.(1992)]{1992ApJ...385..294W} Wood, J.~H., Horne, K., 
\& Vennes, S.\ 1992, \apj, 385, 294 
\bibitem[Wu 
\& Wickramasinghe(1991)]{1991MNRAS.252..386W} Wu, K., \& Wickramasinghe,
D.~T.\ 1991, \mnras, 252, 386 
\bibitem[Wynn et al.(1997)]{1997MNRAS.286..436W} Wynn, G.~A., King, A.~R., 
\& Horne, K.\ 1997, \mnras, 286, 436 
\bibitem[Young \& Schneider(1980)]{1980ApJ...238..955Y} Young, P., \&
  Schneider, D.~P.\ 1980, \apj, 238, 955 
\bibitem[Young et al.(1981)]{1981ApJ...245.1035Y} Young, P., Schneider, 
D.~P., \& Shectman, S.~A.\ 1981, \apj, 245, 1035 
\bibitem[Zhang et al.(1995)]{1995ApJ...454..447Z} Zhang, E., Robinson, 
E.~L., Stiening, R.~F., \& Horne, K.\ 1995, \apj, 454, 447 

\end{thebibliography}
\end{document}